\documentclass[12pt]{article}
\usepackage{amsmath, amsthm, amssymb,latexsym,enumerate}
\usepackage{color}

\newcommand{\be}{\begin{equation}}
\newcommand{\ee}{\end{equation}}
\def\bea{\begin{eqnarray}}
\def\eea{\end{eqnarray}}
\def\bean{\begin{eqnarray*}}
\def\eean{\end{eqnarray*}}
                      
\def\m#1{\mathcal#1}
\def\bea{\begin{eqnarray}}
\def\eea{\end{eqnarray}}
\def\sla{\raise.15ex\hbox{$/$}\kern-.57em}

\renewcommand{\d}{\partial}
\newcommand{\al}{\alpha}

\newcommand{\nn}{\nonumber}

\newcommand{\deltab}{\boldsymbol\delta}
\newcommand{\dst}{\displaystyle}
\newcommand{\ba}{\begin{array}}
\newcommand{\ea}{\end{array}}
\newcommand{\mc}{\mathcal}
\newcommand{\ov}{\overline}
\renewcommand{\phi}{\varphi}
\newcommand{\phibar}{\ov\varphi{}}
\newcommand{\ta}{\theta}
\newcommand{\tabar}{\ov\theta{}}
\newcommand{\chibar}{\ov\chi{}}
\newcommand{\Abar}{\ov{A}{}}
\newcommand{\Cbar}{\ov{C}{}}
\newcommand{\Tbar}{\ov{T}{}}
\newcommand{\om}{\omega}
\newcommand{\ombar}{\ov{\omega}{}}
\newcommand{\dbar}{\ov{d}{}}
\newcommand{\qbar}{\ov{q}{}}
\newcommand{\sbar}{\ov{s}{}}
\newcommand{\p}{\partial}
\newcommand{\half}{\frac{1}{2}}
\newcommand{\qter}{\frac{1}{4}}
\newcommand{\ep}{\epsilon}
\newcommand{\eps}{\varepsilon}
\newcommand{\epsbar}{{\ov\varepsilon}{}}
\newcommand{\epbar}{{\ov\epsilon}{}}
\newcommand{\da}{\delta}
\newcommand{\dab}{\boldsymbol\delta}
\newcommand{\cQ}{\mc{Q}}
\newcommand{\cS}{\mc{S}}
\newcommand{\Qbar}{\ov{\cQ}{}}
\newcommand{\Sbar}{\ov{\cS}{}}
\newcommand{\wh}{\widehat}
\newcommand{\cK}{\mc{K}}

\newcommand{\esum}{\sum_\text{even}}
\newcommand{\osum}{\sum_\text{odd}}
\newcommand{\cW}{\mc{W}}
\newcommand{\Wbar}{\ov\cW{}}
\newcommand{\la}{\lambda}

\begin{document}
\renewcommand{\theequation}{\thesection.\arabic{equation}}

\begin{titlepage}
\begin{flushright}
{\small
 {\tt ULB-TH/09-37}
}
\end{flushright}
\vspace{1cm}

\begin{center}
{\bf \Large The BLG Theory in Light-Cone Superspace}
\end{center}
\vskip 1cm

\begin{center}{
{\bf Dmitry Belyaev${}^{\,1,a}_{}$, Lars Brink${}^{\,2,b}_{}$,  Sung-Soo Kim${}^{\,3,c}_{}$, and Pierre Ramond${}^{1,d}_{}$
}
\vskip .9cm
{\em  ${}^{1~}_{}$Institute for Fundamental Theory,\\
Department of Physics, University of Florida\\
Gainesville FL 32611, USA}
\vskip .2cm

{\em ${}^{2~}_{}$Department of Fundamental Physics \\
Chalmers University of Technology, \\
S-412 96 G\"oteborg, Sweden}
\vskip .2cm

{\em ${}^{3~}_{}$Physique Th\'{e}orique et Math\'{e}matique\\
Universit\'{e} Libre de Bruxelles and International Solvay Institutes,\\
ULB-C.P. 231, B-1050 Bruxelles,
Belgium}
}\end{center}

\vskip .9cm
\begin{abstract}
\noindent  
The light-cone superspace version of the $d=3, N=8$ superconformal theory of Bagger, Lambert and Gustavsson (BLG) is obtained as a solution to constraints imposed by $OSp(2,2|8)$ superalgebra. The Hamiltonian of the theory is shown to be a quadratic form of the dynamical supersymmetry transformation.
\end{abstract}

\vskip 1.4cm

\noindent Keywords: Superspace; Light-cone; Superconformal Theories; Chern-Simons Theories.

\vspace{1cm}
\vfill \vskip 5mm \hrule width 5.cm \vskip 2mm {\small
\noindent $^a~$belyaev@phys.ufl.edu\\
\noindent $^b~$lars.brink@chalmers.se\\
\noindent $^c~$sungsoo.kim@ulb.ac.be\\
\noindent $^d~$ramond@phys.ufl.edu
}
\end{titlepage}

\newpage

\setcounter{equation}{0}
\section{Introduction}
The $d=3, N=8$ superconformal theory has recently been formulated covariantly by Bagger and Lambert~\cite{BAG}{}, and  Gustavsson~\cite{GUS}{}, and its light-cone superspace formulation has been given in~\cite{BENGT, DB}{}.  In this paper, we will report on the use of algebraic techniques to construct this theory in light-cone superspace from its $OSp(\,2,2\,|\,8\,)$ superconformal symmetry, using the same superfield (in one less dimension) that describes $d=4, N=4$ SuperYang-Mills. 

The introduction of supersymmetry into quantum field theory has led to new restrictions in their quantum behavior. These effects are most spectacular in maximally supersymmetric theories. Although seemingly far from the real world, these theories constitute a starting point for the discussion of the r\^ole of symmetries in quantum field theories.  It was realized long ago that the maximally supersymmetric $d=4, N=4$ Yang-Mills theory has unique properties, such as being finite in perturbation theory~\cite{Brink:1982wv}. More recently it has been shown that $d=4, N=8$ supergravity also has remarkable properties in perturbation theory being finite at least up to four loops~\cite{Bern:2009kd}. The underlying symmetry of the Yang-Mills theory  is  the full superconformal symmetry, $PSU(2, 2|4)$, while the symmetry in the supergravity case is the SuperPoincar\' e group times Cremmer and Julia's $E_{7(7)}$~\cite{Cremmer:1979up} symmetry.

In a program that we have followed for quite some time~\cite{Bengtsson:1983pg} we have studied these theories and the corresponding ones in other space-time dimensions in Dirac's light-front form~\cite{Dirac}. In this formalism we only use the physical degrees of freedom and the full SuperPoincar\'e algebra is non-linearly realized. It is the light-cone gauge formalism since we can reach the same result by the gauge choice that a light-cone component of the gauge fields be zero and by use of equations of motion to solve for the remaining unphysical degrees of freedom. We have found in this formalism a great similarity between the two classes of maximally supersymmetric theories and that they are each described by a superspace and a corresponding superfield that are universal.

The first superspace with {\it eight} complex Grassmann variables is used to describe maximally supersymmetric supergravity theories: $N=1$ in $d=11$, $N=8$ in $d=4$, $N=16$ in $d=3$, and so on. With a dimensionful coupling, these theories are not superconformal. They respect instead the non-compact and non-linear symmetries, $E_{7(7)}$ in $d=4$, $E_{8(8)}$ in $d=3$, etc., with light-cone superspace formulation written in terms of the same constrained chiral superfield~\cite{Brink:2008qc,Brink:2008hv}.

The second superspace with {\it four} complex Grassmann variables is equally rich. It houses theories with maximal superconformal symmetry in $d=6,5,4$ and $3$ dimensions, as well as other maximally supersymmetric gauge theories such as $N=1$, $d=10$ SuperYang-Mills. It has already been shown~\cite{ANANTH} how the fully interacting $d=4, N=4$ SuperYang-Mills theory~\cite{N=4} can be determined by requiring $PSU(\,2,2\,|\,4\,)$ superconformal symmetry on a constrained chiral superfield in this light-cone superspace. In this paper, we will present a similar analysis of the $d=3, N=8$ superconformal theory.  This will be an alternative way to find the BLG-theory, which will open up new venues to investigate the model and to find its limitations and possible extensions.

In the light-cone formulation (on the light front), symmetries split into kinematical and dynamical ones. Kinematical symmetries are linearly realized, while dynamical ones contain a linear term (free theory), and terms non-linear in the (super)fields. In superconformal theories, dynamical supersymmetries suffice to {\it completely} determine the theory algebraically. Our technique is to use algebraic consistency to find all possible non-linear realizations of the algebra on the chiral superfields.  

\setcounter{equation}{0}
\section{The $N=4$ Chiral Superfield}
Introduce the usual light-cone variables in $d$ spacetime dimensions, $x^\pm_{}=(x^0_{}\pm\, x^{d-1}_{})/\sqrt{2}$, and their derivatives 
$\p^\pm_{}=(\p^0_{}\pm\, \p^{d-1}_{})/{\sqrt{2}}$ satisfying 
$[\,\p^{+}\,,\, x^{-}\,]\,=\,[\,\p^{-}\,,\,x^{+}\,]\,  =\, -1\ ,$ with the  metric $\eta^{\mu\nu}= (-,+,\cdots,+)$. The $N=4$ superspace contains four complex anticommuting Grassmann variables, $\theta^m$, ($m=1,...,4$) and their conjugates $\tabar_m$, 

Fundamental to this superspace are the chiral superfields  
\begin{align}
\phi^{\,a}_{}\,(y)&=~\frac{1}{ \p^+}\,A^{\,a}_{}\,(y)\,+\,\frac{i}{\sqrt 2}\,{\theta_{}^m}\,{\theta_{}^n}\,{\Cbar^{\,a}_{mn}}\,(y)\,+\,\frac{1}{12}\,{\theta_{}^m}\,{\theta_{}^n}\,{\theta_{}^p}\,{\theta_{}^q}\,{\eps_{mnpq}}\,{\p^+}\,{\Abar}^{\,a}_{}\,(y)\nn\\
& ~~~ +~\frac{i}{\p^+}\,\theta^m_{}\,\chibar^{\,a}_m(y)+\frac{\sqrt 2}{6}\theta^m_{}\,\theta^n_{}\,\theta^p_{}\,\eps^{}_{mnpq}\,\chi^{q\,a}_{}(y) ,\label{comp}
\end{align}
where $a$ is a taxonomic index and $
y=(\,x^{}_1,...,x^{}_{d-2}\,,x^{+},\,x^--i\theta^m\tabar_m/\sqrt{2}\,)$ are chiral coordinates{}\,\footnote{
We will take $x^{+}=0$ for the light-front surface with respect to which the generators of $OSp(2,2|8)$ will be defined. 
}.
The superfields are chiral
\be
d^m_{}\,\phi^a\,(y)~=~0\ ,
\ee
and obey the ``inside-out'' constraint  
\be
\dbar^{}_m\,\dbar^{}_n\,\phi^a~=~\frac{1}{2}\,\eps^{}_{mnpq}\,d^p_{}\,d^q_{}\,\phibar^a\ ,\label{inout}
\ee
where the chiral derivatives
\begin{align}
d^{m}_{}~=~-\frac{\p}{\p\tabar_m}\,-\,\frac{i}{\sqrt2}\,\theta^m\,\p^+\ ;\qquad \dbar_{n}~=~ \frac{\p}{\p\theta_n}\,+\,\frac{i}{\sqrt2}\,\tabar_n\,\p^+\ ,
\end{align}
satisfy 
\be
\{\,d^m_{}\,,\,\dbar_{n}\,\}~=-i\,\sqrt2\,\da^m{}_n\,\p^+\ .
\ee 
The component fields $A,\,\Abar, $ and $\Cbar_{mn}$ represent eight bosons; $\chi^m$ and $\chibar_m$ are the eight fermions.

\setcounter{equation}{0}
\section{The Superconformal Algebra in $d=3$: $OSp(2,2\,|\,8)$ }
In $d=6,5,4$, and $3$, chiral superfields \eqref{comp} form a linear representation of the conformal superalgebras in these dimensions. In $d=4$, they can be used to describe the interacting $N=4$ SuperYang-Mills theory, with a representation of the $PSU(2,2|4)$ superalgebra non-linear in the superfield. The same superfields, in $d=3$, can be used to describe the interacting $N=8$ SuperChern-Simons (BLG) theory, with the superalgebra $OSp(2,2|8)$ realized nonlinearly. This superalgebra has the following bosonic subalgebra
\[
SO(8)\times Sp(2,2) \subset OSp(2,2\,|\,8) \ ,\]
where $SO(8)$ is the $R$-symmetry, and $Sp(2,2)\sim SO(3,2)$ is the conformal group in three dimensions. Below we will give a representation of this superalgebra in terms of operators corresponding to a free (non-interacting) theory. (See also Appendix \ref{App:algebra}.) 

\subsection{$R$-symmetries}
The action of the $R$-symmetry on the chiral superfield is expressed in terms of the operators (the kinematical supersymmetry generators)
\be
q^{m}_{}~=~-\frac{\p}{\p\tabar_m}\,+\,\frac{i}{\sqrt2}\,\theta^m\,\p^+\ ;\quad \qbar_{n}~=~ \frac{\p}{\p\theta^n}\,-\,\frac{i}{\sqrt2}\,\tabar_n\,\p^+\ ,
\ee
which satisfy  
\be
\label{qqbar}
\{\,q^m_{}\,,\,\qbar_{n}\,\}~=~i\,\sqrt2\,\da^m{}_n\,\p^+\ .
\ee
They do not affect chirality since they anticommute with the chiral derivatives. 

The $SO(8)$ $R$-symmetry is written as $SO(6)\times SO(2)\sim SU(4)\times U(1)$ transformations, with generators $T^m{}_n$, and $T$,~
%
\bea
\label{U4}
\da^{}_{SU(4)}\,\phi^a_{}&=&\omega^m{}_n T^n{}_m\phi^{a}~=
~\omega^m{}_n\,\frac{i}{\sqrt{2}}\left( 
q^n_{}\,\qbar_m^{}-\frac{1}{4}\da^n_{\,m}\,q^k_{}\,\qbar_k^{}\,\right)
\frac{1}{\p^+_{}}\,\phi^a_{}\ , \nn\\
\da^{}_{U(1)}\,\phi^a_{}&=& \omega T\phi^{a}~=~
\omega \frac{i}{4\sqrt{2}}\left( q^m_{}\,\qbar_m^{}-\qbar_m^{}\, q^m_{}\,\right)\frac{1}{\p^+_{}}\,\phi^a_{}\ ,
\eea
together with the coset transformations, with generators  $T^{m n}$, and $\Tbar_{m n}$,
\begin{align}
\label{coset}
\da^{}_{\ov{coset}}\,\phi^a_{}~&=~\omega^{mn}\Tbar_{mn}\phi^{a}~=
~\omega^{mn}_{}\,\frac{i}{\sqrt{2}}\, \qbar^{}_m\,\qbar_n^{}\,\frac{1}{\p^+_{}}\,\phi^a_{}\,
 ; \nn\\
\da^{}_{coset}\,\phi^a_{}~&=~\ombar_{mn}T^{mn}\phi^{a}~=~
{\ombar}^{}_{mn}\,\frac{i}{\sqrt{2}}\,  q^{m}_{}\, q_{}^n\,\frac{1}{\p^+_{}}\,\phi^a_{}\ , 
\end{align}
completing the full $SO(8) \supset SO(6)\times SO(2)$. 
All $R$-symmetry generators are kinematical.

\subsection{Superconformal Symmetries}
Space-time generators are either kinematical or dynamical. Kinematical generators operate within the initial surface, 
 while the dynamical ones act transversely to the initial surface, and define the dynamics. The kinematical generators are the same in free and interacting theories, and induce changes linear in the fields. The dynamical generators, on the other hand, contain a part linear in the (super)fields for the free theory, as well as terms which are non-linear in the (super)fields, accounting for the interactions. 

In light-cone notation, the ten generators of the conformal group in three dimensions are given by  
\bean
&&~~{\rm  Lorentz~ Group: }\quad J^{+-}\ , J^+ \ ;\quad \mc{J}^-\\
&&~~{\rm  Translations: }\qquad P\ , P^+\ ;\quad \mc{P}^-\\
&&~~{\rm  Dilatation: }\qquad\quad D \\
&&~~{\rm  Conformal: }\qquad\quad  K\ ,  K^+\ ;\quad \mc{K}^-
\eean
with the dynamical generators written in calligraphic letters. Note that $J^{+-}$ and $K^+$, $K$ and $D$ are kinematical \emph{only} at $x^+=0$ (cf.~\cite{Bengtsson:1983pg}). The supersymmetry and superconformal (or conformal supersymmetry) generators, which complete the superconformal algebra, also split into kinematical and dynamical generators
\bean
&&~~{\rm  Supersymmetry: }\quad q^m\ , \qbar_m\ ;\quad \cQ^m\ , \Qbar_m\\
&&~~{\rm  Superconformal: }\quad s^m\ , \sbar_m\ ;\quad \cS^m\ , \Sbar_m\ .
\eean

\subsection{Kinematical Transformations} 
The kinematical conformal group transformations are given by 
\be
\da^{}_{P^{+}}\,\phi^a_{}~=-i\,\p^+_{}\,\phi^a_{}\ ;\qquad 
\da^{}_{P}\,\phi^a_{}~=-i\,\p\,\phi^a_{}\ ; \nn
\ee
\be
\da^{}_{J^{+}}\,\phi^a_{}~=~i x\,\p^+_{}\,\phi^a_{}\ ;\qquad 
\da^{}_{J^{+-}}\,\phi^a_{}~=~i(\, \mc{A}+\frac{x}{2}\p+\half \,)\,\phi^a_{}\ ; \nn
\ee
\be
\da^{}_{D}\,\phi^a_{}~=~i\,(\,\mc A\,-\,\frac{x}{2}\,\p\,)\,\phi^a_{}\ ;\quad
\da^{}_{K}\,\phi^a_{}~=~2i\,x\,\mc A\,\phi^a_{} \ ;\quad
\da^{}_{K^+}\,\phi^a_{}~=i\,x^2\p^+_{}\,\phi^a_{}\ ,
\label{kinconf}
\ee
where $\p$ is the derivative with respect to the lone transverse variable $x$ in the superfield, and
\be
\label{defA}
{\mc A}~\equiv~x^-\,\p^+-\frac{x}{2}\,\p\,-\frac{1}{2}{\mc N}+\frac{1}{2}\ ;\qquad{}
{\mc N}~\equiv~\theta^m_{}\frac{\p}{\p\theta^m_{}}\,+\,\tabar^{}_m\frac{\p}{\p\tabar^{}_m}\ .
\ee
The kinematical (spectrum-generating) supersymmetries are  
\be
\da^{\,}_{\eps\qbar}\,\phi^a_{}~=~\eps^m_{}\qbar_m^{}\,\phi^a_{}\ ; \qquad \da^{\,}_{\epsbar q}\,\phi^a_{}~=~\epsbar^{}_m q^m_{}\,\phi^a_{}\ ,
\ee
and the kinematical superconformal transformations are
\be
\da^{}_{\eps\sbar}\,\phi^a_{}~=-i x\, \eps^m\,\qbar_m^{}\,\phi^a_{}\ ;\qquad
\da^{}_{\epsbar s}\,\phi^a_{}~=~i x\, \epsbar_m\,q^m_{}\,\phi^a_{}\ .
\ee
where $\eps^m$ and $\epsbar_m$ are anticommuting parameters.

\subsection{Free Dynamical Transformations}
A distinguishing feature of superconformal theories is that {\it all} dynamical generators are determined by commutations from the dynamical supersymmetry generators. Starting from the {\em free} dynamical supersymmetry transformations{}\,\footnote{
For emphasis, we write \emph{dynamical} transformations with a bold $\dab$.
}, 
\be
\label{freeds}
\dab^{free}_{\eps\Qbar}\,\phi^a~=~\frac{1}{\sqrt{2}}\eps_{}^m\qbar^{}_m\,\frac{\p}{\p^+_{}}\,\phi^a_{}\ ,\qquad 
\dab^{free}_{\epsbar\cQ}\,\phi^a~=~\frac{1}{\sqrt{2}}\epsbar^{}_m q^{m}_{}\,\frac{\p}{\p^+_{}}\,\phi^a_{}\ ,\ee
we use the algebra 
\bea
\label{sven}
\ba[b]{rclcl}
[\,\dab_{\eps\Qbar}\,,\,\dab_{\epsbar\cQ}\,]\,\phi^a &=&  \sqrt2\,\epsbar_m\eps^m\dab_{{\mc P}^{-}}\,\phi^a
&\quad\rightarrow\quad& \dab_{\mc{P}^{-}}\,\phi^a\ , \\[5pt]
[\,\da_K\,,\,\dab_{\mc{P}^{-}}\,]\,\phi^a &=& 2i\,\dab_{\mc{J}^{-}}\,\phi^a 
&\rightarrow& \dab_{\mc{J}^{-}}\,\phi^a\ , \\[5pt]
[\,\da_K\,,\,\dab_{\mc{J}^{-}}\,]\,\phi^a &=& -i\,\dab_{\mc{K}^{-}}\,\phi^a
&\rightarrow& \dab_{\mc{K}^{-}}\,\phi^a\ , \\[5pt]
[\,\da_K\,,\,\dab_{\eps\Qbar}\,]\,\phi^a &=& \sqrt2\,\dab_{\eps\Sbar}\,\phi^a 
&\rightarrow& \dab_{\eps\Sbar}\,\phi^a\ , \\[5pt]
[\,\da_K\,,\,\dab_{\epsbar\cQ}\,]\,\phi^a &=& -\sqrt2\,\dab_{\epsbar\cS}\,\phi^a 
&\rightarrow& \dab_{\epsbar\cS}\,\phi^a\ ,
\ea
\eea
to obtain the remaining dynamical transformations, 
\bea
\label{dyngen}
\ba[b]{rcrcl}
\text{``{\rm Time}''}~(x^{+})~{\rm Translation:}&&\dab^{free}_{\mc{P}^{-}_{}}\,\phi^{a}_{}
&=&\dst -i\,\frac{\p^2_{}}{2\,\p^+}\,\phi^a_{}\ ,
\\[8pt] 
{\rm Lorentz~ Boost:}&&\dab^{free}_{\mc{J}^{-}}\,\phi^{a}_{}
&=&\dst -i\frac{\p}{\p^+}{\mc A}\,\phi^a_{} \ ,
\\[8pt]
{\rm Conformal~ Boost:}&&\dab^{free}_{\mc{K}^{-}}\,\phi^{a}_{}
&=&\dst 2i\,\frac{1}{\p^+}\,{\mc A}\,({\mc A}-\frac{1}{2})\,\phi^a_{} \ ,
\\[8pt]
{\rm Superconformal:}&&\dab^{free}_{\eps\Sbar}\,\phi^a
&=&\dst i\,\eps^m\qbar^{}_m\,\frac{1}{\p^+_{}}\,\mc{A}\,\phi^a_{}\ ,
\\[5pt] 
&&\dab^{free}_{ \epsbar\cS}\,\phi^a
&=&\dst -i\,\epsbar_m q^m_{}\,\frac{1}{\p^+_{}}\,\mc{A}\,\phi^a_{}\ .
\ea
\eea
This representation of the dynamical generators is valid in the free theory, and needs to be augmented in the interacting theory. Together with the kinematical generators, they satisfy the $OSp(2,2\,|\,8)$ algebra, whose light-cone commutation relations appear in Appendix \ref{App:algebra}.

\setcounter{equation}{0}
\section{Interactions}
In the interacting theory, the dynamical generators acquire contributions nonlinear in the superfields. To specify the full theory, we need only find these contributions to the dynamical supersymmetry generators. All other dynamical generators follow from the algebra by commutations. 

\subsection{Kinematical Constraints}\label{sec:4.1}
The dynamical supersymmetries consist of two parts: 

\be
\deltab^{}_{\varepsilon\overline {\mathcal Q}}\,\varphi^a~=~\deltab^{ free}_{\varepsilon\overline {\mathcal Q}}\,\varphi^a+\deltab^{int}_{\varepsilon\overline {\mathcal Q}}\,\varphi^a\ ,\qquad \deltab^{}_{\overline\varepsilon {\mathcal Q}}\,\varphi^a~=~\deltab^{ free}_{\overline\varepsilon {\mathcal Q}}\,\varphi^a+\deltab^{ int}_{\overline\varepsilon {\mathcal Q}}\,\varphi^a\ .
\ee
The forms of $\deltab^{int}_{\varepsilon\overline {\mathcal Q}}\,\varphi^a$ and $\deltab^{ int}_{\overline\varepsilon {\mathcal Q}}\,\varphi^a$ are  highly restricted by the following ten algebraic constraints~\cite{PRtalk}:

\begin{enumerate} [(i)]

\item \label{1} Chirality: the transformations should be chiral, that is,  

\be
d^m\,(\deltab^{int}_{\varepsilon\overline {\mathcal Q}}\,\varphi^a)~=~d^m\,(\deltab^{int}_{\overline\varepsilon {\mathcal Q}}\,\varphi^a)~=~0\ ,\ee
and satisfy the inside-out constraint,

\bea
\label{IOC}
\dab_{\epsbar\cQ}^{int}\phi^a=\frac{d^{[4]}}{2\p^{+2}}\left( \dab_{\ep\Qbar}^{int}\phi^a \right)^\ast ,
\eea
where $
d^{[4]}\equiv d^1 d^2 d^3 d^4$.

\item \label{2} Both are independent of $x^-$, since

\be
[\,\delta^{}_{P^+}\,,\,\deltab^{}_{\varepsilon\overline {\mathcal Q}}\,]\,\varphi^a~=~[\,\delta^{}_{P^+}\,,\,\deltab^{}_{\overline\varepsilon {\mathcal Q}}\,]\,\varphi^a~=~0
\ .\ee

\item \label{3} Both are also independent of $x$, as

\be
[\,\delta^{}_{P}\,,\,\deltab^{}_{\varepsilon\overline {\mathcal Q}}\,]\,\varphi^a~=~[\,\delta^{}_{P}\,,\,\deltab^{}_{\overline\varepsilon {\mathcal Q}}\,]\,\varphi^a~=~0\ .
\ee

\item \label{4} Neither have transverse derivatives $\partial$: from

\be
[\,\delta^{}_{J^+}\,,\,\deltab^{}_{\overline\varepsilon {\mathcal Q}}\,]\,\varphi^a~=~\frac{i}{\sqrt{2}}\,\delta^{}_{\bar\varepsilon q}\,\varphi^a\ ,\quad [\,\delta^{}_{J^+}\,,\,\deltab^{}_{\varepsilon\overline {\mathcal Q}}\,]\,\varphi^a~=~\frac{i}{\sqrt{2}}\,\delta^{}_{\varepsilon\bar q}\,\varphi^a\ ,
\ee
it follows that

\be
[\,\delta^{}_{J^+}\,,\,\deltab^{int}_{\overline\varepsilon {\mathcal Q}}\,]\,\varphi^a~=~[\,\delta^{}_{J^+}\,,\,\deltab^{int}_{\varepsilon\overline {\mathcal Q}}\,]\,\varphi^a~=~0\ .\ee

\item \label{5} From 

\be
[\,\delta^{}_{\bar\varepsilon q}\,,\,\deltab^{}_{\varepsilon\overline {\mathcal Q}}\,]\,\varphi^a~=-\bar\varepsilon_m\varepsilon^m\,\delta^{}_P\,\varphi^a_{}\ ,\qquad [\,\delta^{}_{\varepsilon \bar q}\,,\,\deltab^{}_{\overline\varepsilon {\mathcal Q}}\,]\,\varphi^a~=~\bar\varepsilon_m\varepsilon^m\,\delta^{}_P\,\varphi^a_{}
\ ,\ee
we deduce that 

\be 
[\,\delta^{}_{\bar\varepsilon q}\,,\,\deltab^{int}_{\varepsilon\overline {\mathcal Q}}\,]\,\varphi^a~=~[\,\delta^{}_{\varepsilon \bar q}\,,\,\deltab^{int}_{\overline\varepsilon {\mathcal Q}}\,]\,\varphi^a~=~0\ .
\ee

\item \label{6} Proper transformation under $J^{+-}$ require

\be
[\,\delta^{}_{J^{+-}}\,,\,\deltab^{int}_{\varepsilon\overline {\mathcal Q}}\,]\,\varphi^a~=~\frac{i}{2}\,\deltab^{int}_{\varepsilon\overline {\mathcal Q}}\,\varphi^a\ ,
\qquad [\,\delta^{}_{J^{+-}}\,,\,\deltab^{int}_{\overline\varepsilon {\mathcal Q}}\,]\,\varphi^a~=~\frac{i}{2}\,\deltab^{int}_{\overline\varepsilon {\mathcal Q}}\,\varphi^a.
\ee

\item \label{7} Proper transformations under D require
\be
[\,\delta^{}_{D}\,,\,\deltab^{int}_{\varepsilon\overline {\mathcal Q}}\,]\,\varphi^a~=-\frac{i}{2}\,\deltab^{int}_{\varepsilon\overline {\mathcal Q}}\,\varphi^a\ ,\qquad [\,\delta^{}_{D}\,,\,\deltab^{int}_{\overline\varepsilon {\mathcal Q}}\,]\,\varphi^a~=-\frac{i}{2}\,\deltab^{int}_{\overline\varepsilon {\mathcal Q}}\,\varphi^a\ .
\ee

\item \label{8} They have opposite $U(1)$ $R$-charge,

\be
[\,\delta^{}_{U(1)}\,,\,\deltab^{int}_{\varepsilon\overline {\mathcal Q}}\,]\,\varphi^a~=-\frac{1}{2}\,\deltab^{int}_{\varepsilon\overline {\mathcal Q}}\,\varphi^a\ ,\qquad [\,\delta^{}_{U(1)}\,,\,\deltab^{int}_{\overline\varepsilon {\mathcal Q}}\,]\,\varphi^a~=~\frac{1}{2}\,\deltab^{int}_{\overline\varepsilon {\mathcal Q}}\,\varphi^a\ .
\ee

\item \label{9} The eight interacting supersymmetries must also transform as an $SO(8)$ vector, that is,  with $\bar\varepsilon'^{}_m=2\overline\omega^{}_{mn}\varepsilon^n_{}$,

\be\label{cosetbar}
[\,\delta^{}_{\overline{coset}}\,,\,\deltab^{}_{\varepsilon\overline {\mathcal Q}}\,]\,\varphi^a~=~0\ , 
\qquad 
[\,\delta^{}_{{coset}}\,,\,\deltab^{}_{\varepsilon\overline {\mathcal Q}}\,]\,\varphi^a~=~\deltab^{}_{ \bar\varepsilon'{\mathcal Q}}\,\varphi^a\ .
\ee
Similarly, with $\varepsilon'^m=2\omega^{mn}\overline\varepsilon_n$,

\be\label{cosetbar2}
[\,\delta^{}_{\overline{coset}}\,,\,\deltab^{}_{\overline\varepsilon {\mathcal Q}}\,]\,\varphi^a~=~\deltab^{}_{ \varepsilon'\overline{\mathcal Q}}\,\varphi^a\ , 
\qquad 
[\,\delta^{}_{{coset}}\,,\,\deltab^{}_{\overline\varepsilon {\mathcal Q}}\,]\,\varphi^a~=~0\ .
\ee

\item \label{10} Both $\deltab^{\,\rm int}_{\varepsilon\overline {\mathcal Q}}\,\varphi^a$ and $\deltab^{int}_{\overline\varepsilon {\mathcal Q}}\,\varphi^a$ are cubic powers of the superfields. 

In three dimensions, canonical Bose fields have mass dimension of one-half, so that the chiral superfield has half-odd integer canonical dimension.  Since we are looking for a conformal theory with no dimensionful parameters, $\deltab^{int}_{\varepsilon\overline{\mathcal Q} }\,\varphi^a$ and $\deltab^{int}_{\overline\varepsilon{\mathcal Q} }\,\varphi^a$ must then both be odd powers of superfields, assuming integer power of derivatives. To allow for three or more superfields, the theory must contain a tensor with at least four indices, $f^{\,a}_{~\,bcd}$ \footnote{In $d=4$, similar considerations suggested a tensor with three indices, $f^{\,a}_{~\,bc}$, which turned out to be the structure functions of the gauge algebra.}.

To see that it is only cubic, we form the combination

\be
\Delta~\equiv~J^{+-}-D~=~i\left(x\partial+\frac12\right)\ ,\ee
where $x\p$ counts the number of transverse variables, and the constant counts the number of superfields. Since $\dab_{\ep\Qbar}^{int}\phi^a$ does not contain any explicit transverse variables, and assuming that it contains products of $n_\phi$ superfields, it follows that 

\bea
[\da_\Delta,\dab_{\ep\Qbar}^{int}]\phi^a=\frac{i}{2}(n_\phi-1)\dab_{\ep\Qbar}^{int}\phi^a\ .
\eea
On the other hand, the algebra requires 
  
\bea
\label{kcDelta}
[\da_\Delta,\dab_{\ep\Qbar}^{int}]\phi^a=i\dab_{\ep\Qbar}^{int}\phi^a,
\eea
where $\da_\Delta\phi^a=\Delta\phi^a$. These agree when $n_\phi=3$, limiting the interacting supersymmetry to a cubic form.

\end{enumerate}
These ten requirements limit the possible forms of the dynamical supersymmetries.

\subsection{Chiral Engineering}
The construction of  chiral polynomials in the superfields is facilitated by the introduction of the coherent state operators \cite{Brink:2008qc,Brink:2008hv}

\be 
E_{\eta}^{}~=~e^{\,\eta\cdot\widehat{\overline d}}_{}\ ,
\ee
where the hat denotes division by $\partial^+$, $\widehat{\overline d}_{m} \equiv \bar d_{m}/\d^{+}$, and $\eta^m$ are arbitrary Grassmann parameters. Since  

\be 
d^m_{}\,\left(\,E_{\eta}^{}\,\varphi^a\,\right)~=~ i\sqrt{2}\,\eta^m_{}\,\left(\,E_{\eta}^{}\,\varphi^a \,\right)\ ,
\ee
$E_{\eta}\,\varphi^a$ are eigenstates of the chiral derivatives. It follows that the quadratic combination

\be
(E^{}_{\eta}\partial^{+B}_{}\,\varphi^b_{})\,(E^{}_{-\eta} \partial^{+C}\,\varphi^c)
\ ,
\ee
is manifestly chiral.
The {\em nested form} 

\be
(E^{}_{\eta}\partial^{+B}_{}\,\varphi^b_{}\,)
\,E^{}_{-\eta} \frac{1}{\partial^{+M}}\left(\,(E^{}_{\zeta}\partial^{+C}\,\varphi^c\,)
(\,E^{}_{-\zeta}\partial^{+D}\,\varphi^d\,)\,\right)\ ,
\ee
is also chiral, and can be used to generate chiral cubic polynomials in the superfields, the coefficients in the series expansion in the independent Grassmann parameters $\eta$ and $\zeta$.

\subsection{Even and Odd Ans\"atze}

To construct the interaction part of the dynamical supersymmetry, we introduce the supersymmetry parameters in the nested Ansatz through the combinations 

\be \label{nested}
E_{\varepsilon}^{}~=~e^{\,\varepsilon\cdot\widehat{\overline q}}_{}\ ,\qquad  E_{\,\bar\varepsilon}^{}~=~e^{\,\bar\varepsilon\cdot{\widehat q}}\ , \ee
which naturally allow to satisfy requirement (\ref{5}), without affecting chirality. This leads us to write the dynamical supersymmetries as a sum of nested ans\" atze of the form 

\bea\label{calK}
\deltab^{\,int}_{\varepsilon\overline {\mathcal Q}}\,\varphi^a&\sim&\frac{f^{\,a}_{~\,bcd}}{\partial^{+A_\alpha}}\left((E_{\varepsilon}^{}E_{\eta}^{}\partial^{+B_\alpha}_{}\varphi^b_{})E_{-\varepsilon}^{}E_{-\eta}^{} \frac{1}{\partial^{+M_\alpha}}\left((E_{\zeta}^{}\partial^{+C_\alpha}\varphi^c)(E_{-\zeta}^{}\partial^{+D_\alpha}\varphi^d\,)\right)\right)\ ,\nonumber\\
&\equiv&
{\cal K}^{a\,(\varepsilon,\eta,\zeta)}_\alpha\ ,
\eea
keeping only the first order in the supersymmetry parameters $\varepsilon^m$. We have to allow for a non-trivial sum over $\alpha$. $f^{\,a}_{~\,bcd}$ and the exponents $A_\alpha$, $B_\alpha$, $M_\alpha$, $C_\alpha$, $D_\alpha$ have yet to be determined. It is convenient to introduce {\em insertion operators} $\mc U_i$ ($i=1,2,3,4$), whose action is defined by  

\bea
{\cal K}^{a\,(\varepsilon,\eta,\zeta)}_\alpha&=&(E_\varepsilon \mc U_1)(E_{-\varepsilon}\mc U_2)\,{\cal K}^{a\,(0,\eta,\zeta)}_\alpha=(E_\zeta \mc U_3)(E_{-\zeta}\mc U_4)\,{\cal K}^{a\,(\varepsilon,\eta,0)}_\alpha \nn \\ &=&( E_\ep E_\eta , E_{-\ep}E_{-\eta} ( E_\zeta ,
E_{-\zeta} ) )\mc{K}_\alpha^{a\,(0,0,0) }
.
\eea
We will often use the useful $(\ ,\ (\ ,\ ))$ notation when we have an operator which makes multiple insertions. 

For this Ansatz, we find that (see Appendix \ref{App:useful} for more details)

\begin{itemize}

\item Chirality (\ref{1}) is manifest
since the $\bar q_n$ anticommute with the chiral derivatives. The inside-out constraint (\ref{IOC}) will be checked below.

\item  (\ref{2}), (\ref{3}), (\ref{4}), and (\ref{5}) are clearly satisfied. 

\item The proper transformation under $J^{+-}$, (\ref{6}), together with the $U(1)$ condition, (\ref{8}), restricts the number of  $\partial^+$ derivatives to four,   

\be \label{ABMCD}
-A_\alpha^{}+B^{}_\alpha-M^{}_\alpha+C^{}_\alpha+D^{}_\alpha~=~4\ ,\ee
which reproduces the correct dimension.

\item The correct $U(1)$ $R$-charge, (\ref{8}), requires after some computation

\be\label{nineb}
\left(\,\eta^m\frac{\partial}{\partial\eta^m}+\zeta^m\frac{\partial}{\partial\zeta^m}~-~4\,\right)\,{\cal K}^{a\,(\varepsilon,\eta, \zeta)}_\alpha~=~0 \ ,
\ee
so that  only the coefficients of the terms \emph{quartic} in $\eta$ and $\zeta$, 

\bea
\label{quartic}
\eta^4, \quad \eta^3\zeta, \quad \eta^2\zeta^2, \quad \eta\zeta^3, \quad \zeta^4
\eea
need to be considered.

\item
We find 
\bea
\label{TbarQbar}
[\da_{\ov {coset}},\dab_{\eps\Qbar}^{int}]\phi^a &=& \frac{1}{\sqrt2}\om^{m n}\sum\left(
\frac{\p}{\p\eta^m}\frac{\p}{\p\eta^n}\mc{S}+\frac{\p}{\p\zeta^m}\frac{\p}{\p\zeta^n}\mc{T}\right)
{\cal K}_\al^{a\,(\ep,\eta,\zeta)},
\eea
where $\mc{S}$ and $\mc{T}$ are multiple insertion operators defined by 

\bea
\label{defST}
\mc{S}\equiv\frac{1}{\p^{+}}(\p^{+},\p^{+}(1,1)), \quad
\mc{T}\equiv(1,\frac{1}{\p^{+}}(\p^{+},\p^{+}))\ .
\eea
The right hand side has to vanish. Because of the appearance of \emph{double} $\eta$- and $\zeta$-derivatives in this expression, the ``even'' and ``odd'' sets,
\bea
(\eta^4, \;\; \eta^2\zeta^2, \;\; \zeta^4) \quad \text{and} \quad
(\eta^3\zeta, \;\; \eta\zeta^3)\ ,
\eea
do not mix. This splits our Ansatz into two:

\begin{itemize}
\item[$\blacktriangleright$]
the \underline{\it even Ansatz}~\footnote{
The nested form (\ref{nested}) was inspired by the structure of the $O(\kappa^2)$ part of the dynamical supersymmetry in $d=3$, $N=16$ supergravity, as given in equation (4.14) in \cite{Brink:2008hv} (the right hand side of that equation should include an extra factor $(-1)^c/(4+2c)!$). Note that our even Ansatz (\ref{evenan}) is its direct analog.
}
\bea
\label{evenan}
\da_{\eps\Qbar}^{int,\, \rm even}\phi^a=\frac{1}{\sqrt2}\esum {\cal K}_\al^{a\,(\eps,\eta,\zeta)}
\Big|_{\eta=\zeta=0;\text{ linear in $\eps$}},
\eea 
where the sum stands for the operator
\bea
\label{esum}
\esum &\equiv& \sum_{\al=0,\pm1} 
(-1)^\al\frac{\p}{\p\eta^{[2-2\al]}}\frac{\p}{\p\zeta^{[2+2\al]}}
\eea
with
\bea
\frac{\p}{\p\eta^{[2-2\al]}}\frac{\p}{\p\zeta^{[2+2\al]}} &\equiv& 
\frac{\ep^{i_1\dots i_4}}{(2-2\al)!(2+2\al)!}
\frac{\p}{\p\eta^{i_1\dots i_{2-2\al}}}\frac{\p}{\p\zeta^{i_{3-2\al}\dots i_4}}
\eea
\item[$\blacktriangleright$]
and the \underline{\it odd Ansatz} 
\bea
\label{oddan}
\da_{\eps\Qbar}^{int,\, \rm odd}\phi^a=\frac{1}{\sqrt2}\osum {\cal K}_\al^{a\,(\eps,\eta,\zeta)}
\Big|_{\eta=\zeta=0;\text{ linear in $\eps$}},
\eea 
where
\bea
\label{osum}
\osum &\equiv& \sum_{\al=\pm1/2} 
(-1)^{\al+\half}\frac{\p}{\p\eta^{[2-2\al]}}\frac{\p}{\p\zeta^{[2+2\al]}}
\eea
\end{itemize}
We then find that the first  constraint in (\ref{cosetbar}) is satisfied for both ans\"atze provided
\bea
\label{rec1}
\cK_{\al+1}^a=\mc{S}^{-1}\mc{T}\cK_\al^a=\p^{+}\left(\frac{1}{\p^{+}},\frac{1}{\p^{+2}}(\p^{+},\p^{+})\right)\cK_\al^a,
\eea
which yields a recursion relation for the powers of $\p^{+}$
\bea
\label{rec2}
&& A_{\al+1}=A_\al-1, \quad B_{\al+1}=B_\al-1, \quad M_{\al+1}=M_\al+2 \nn\\
&& C_{\al+1}=C_\al+1, \quad D_{\al+1}=D_\al+1.
\eea
\item
To verify the second  constraint  in (\ref{cosetbar}), we find that
\bea
\label{TQbar}
[\da_{coset},\dab_{\eps\Qbar}^{int}]\phi^a &=&2 i\ombar_{m n}\eps^n\sum\eta^m\mc{S}^{-1}\cK_\al^{a\,(0,\eta,\zeta)} \nn\\[5pt]
&& -i\ombar_{m n}\sum\left(\eta^m\eta^n\mc{S}^{-1}+\zeta^m\zeta^n\mc{T}^{-1}\right)\cK_\al^{a\,(\ep,\eta,\zeta)}\ .
\eea
 The sum in the second line vanishes in both the even and odd cases, thanks to the recursion relation (\ref{rec1}). The second constraint in (\ref{cosetbar}) is then satisfied provided
\bea
\label{ansatzbar}
\dab_{\epsbar\cQ}^{int}\phi^a=i\epsbar_m\sum\eta^m\mc{S}^{-1}\cK_\al^{a\,(0,\eta,\zeta)}
=\frac{1}{\sqrt2}\sum\cK_\al^{a\,(\epsbar,\eta,\zeta)}\Big|_\text{linear in $\epbar$}\ .
\eea
We have verified that the inside-out constraint (\ref{IOC}) is in agreement with (\ref{ansatzbar}) in both the even and odd case, thanks again to the recursion relation (\ref{rec2}). (See Appendix~\ref{App:useful} for more details.)
\end{itemize}
To summarize, we have found that both the even Ansatz (\ref{evenan}) and the odd Ansatz (\ref{oddan}), with the exponents $A_\al$, $B_\al$, $M_\al$, $C_\al$, $D_\al$ satisfying the dimensional constraint (\ref{ABMCD}) and the recursion relation (\ref{rec2}), are solutions to the chirality, inside-out and all the kinematical constraints. 
Next, we turn to satisfying the dynamical constraints.

\subsection{$\dab_{\mc{P}^{-}}\phi^a$,  $\dab_{\eps\Sbar}\phi^a$, and $\dab_{\mc{K}^{-}}\phi^a$} 
Having found ans\"atze for $\dab_{\eps\Qbar}^{int}\phi^a$ and $\dab_{\epsbar\cQ}^{int}\phi^a$ that satisfy all the kinematical constraints, we can use (\ref{sven}) to calculate the remaining dynamical transformations which will automatically satisfy their own kinematical constraints thanks to the Jacobi identities. There is a subtlety, however, in the calculation of the Hamiltonian shift $\dab_{\mc{P}^{-}}\phi^a$ via
\bea
\label{QQisP}
[\dab_{\eps\Qbar}^{free}+\dab_{\eps\Qbar}^{int},\dab_{\epsbar\cQ}^{free}+\dab_{\epsbar\cQ}^{int}]\phi^a
=\sqrt2\epbar_m\eps^m\dab_{\mc{P}^{-}}\phi^a,
\eea
as one should verify that the ``off-diagonal'' terms $\epsbar_m\eps^n$, with $m\neq n$, all cancel. The interaction part of the dynamical supersymmetry is linear in $f^a{}_{b c d}$, while the Hamiltonian shift has both linear, $\dab_{\mc{P}^{-}}^{(1)}\phi^a$, and quadratic, $\dab_{\mc{P}^{-}}^{(2)}\phi^a$, parts:
\bea
\dab_{\mc{P}^{-}}^{int}\phi^a=\dab_{\mc{P}^{-}}^{(1)}\phi^a+\dab_{\mc{P}^{-}}^{(2)}\phi^a.
\eea
The $O(f)$ part is then determined by
\bea
\label{QQisP1}
[\dab_{\eps\Qbar}^{free},\dab_{\epsbar\cQ}^{int}]\phi^a+[\dab_{\eps\Qbar}^{int},\dab_{\epsbar\cQ}^{free}]\phi^a
=\sqrt2\epbar_m\eps^m\dab_{\mc{P}^{-}}^{(1)}\phi^a,
\eea
and we verified (see Appendix~\ref{App:P-J-}) that the off-diagonal terms at this order cancel for both the even and odd ans\"atze. In the odd case, the result for the Hamiltonian shift is 

\bea
\label{P-odd}
\dab_{\mc{P}^{-}}^{(1)\,\rm odd}\phi^a=-\frac{i}{2}\frac{\p}{\p r}\left(
\osum K^{a\,[r,1]}_\alpha+\esum K^{a\,[1,r]}_{\al+\half})\right)\Big|_{\eta=\zeta=r=0},
\eea
where 

\be\label{compact}
K^{a\,[r,1]}_\alpha\equiv(E^{}_r\,\mc U^{}_1)(E^{}_{-r}\,\mc U_2)\,{\cal K}^{a\,(0,\eta,\zeta)}_\alpha\ ,
\quad K^{a\,[1,r]}_\alpha\equiv(E^{}_r\,\mc U^{}_3)(E^{}_{-r}\,\mc U_4)\,{\cal K}^{a\,(0,\eta,\zeta)}_\alpha\ ,\ee
introducing the $\p$-exponential,
\bea
E_r=e^{r\wh\p},
\eea
where $\wh\p=\p/\p^{+}$ ($\p$ is the transverse derivative) and $r$ is a dimensionless parameter. The result in the even case is similar and can be obtained from (\ref{P-odd}) by the following substitutions
\bea
\label{oddtoeven}
\osum \;\rightarrow\; \esum, \quad
\esum \;\rightarrow\; -\osum.
\eea

The dynamical superconformal transformations are easily computed, using the transverse  kinematical conformal operator $K$, which acts  as a ladder operator,

\bea
[\da_K,\dab_{\eps\Qbar}]\phi^a=\sqrt2\dab^{}_{\eps\Sbar}\phi^a, \quad{}
[\da_K,\dab_{\epsbar\cQ}]\phi^a=-\sqrt2\dab^{}_{\epsbar\cS}\phi^a\ ,
\eea
which yields, using  $K=2i x\mc{A}$, 

\bea
\dab_{\epsbar\mc{S}}^{int}\phi^a=~\frac{i}{\sqrt2}x\dab_{\epsbar\mc{Q}}^{int}\phi^a\ ,\qquad \dab_{\eps\Sbar}^{int}\phi^a=-\frac{i}{\sqrt2}x\dab_{\eps\Qbar}^{int}\phi^a\ .
\eea
Then  $\dab_{\mc{K}^{-}}\phi^a$ follows from 

\bea
\label{SSisK}
[\dab_{\eps\Sbar},\dab^{}_{\epsbar\cS}]\phi^a=-\frac{1}{\sqrt2}\epsbar_m\eps^m\dab_{\mc{K}^{-}}\phi^a\ .
\eea
The cancellation of its off-diagonal terms follow from 
\bea
[\dab_{\eps\Sbar},\dab^{}_{\epsbar\cS}]\phi^a=-\qter[\da_K,[\da_K,[\dab_{\eps\Qbar},\dab_{\epsbar\cQ}]]]\phi^a\ ,
\eea
which is the  result of   the Jacobi identity JAC$(\da_K,\dab_{\ep\Qbar},\dab^{}_{\epbar\cQ})$, where
\bea
{\rm JAC}(\da_1,\da_2,\da_3)\,:\quad{} 
[\da_1,[\da_2,\da_3]]\phi^a+[\da_2,[\da_3,\da_1]]\phi^a+[\da_3,[\da_1,\da_2]]\phi^a=0\ ,
\eea
 the Jacobi identities JAC$(\da_K,\dab_{\ep\Sbar},\dab^{}_{\epbar\cQ})$ and JAC$(\da_K,\dab_{\ep\Qbar},\dab^{}_{\epbar\cS})$ and the commutation relations

\bea
\label{KSzero2}
[\da_K,\dab_{\eps\Sbar}]\phi^a=0, \quad{}
[\da_K,\dab_{\epsbar\cS}]\phi^a=0\ ,
\eea
whose validity is easily established. Similarly, we find that $[\da_K,\dab_{\mc{K}^{-}}]\phi^a=0$, since
\bea
[\da_K,[\da_K,[\da_K,\dab_{\mc{P}^{-}}]]]=0\ ,
\eea
follows from (\ref{SSisK}), JAC$(\da_K,\dab_{\eps\Sbar},\dab^{}_{\epsbar\cS})$ and (\ref{KSzero2}). The explicit expression for $\dab_{\mc{K}^{-}}\phi^a$ will not be needed in the remainder of our analysis.

\subsection{Dynamical Constraints} 
By definition, the dynamical constraints are the commutation relations of the dynamical transformations from the following complete set,
\bea
\dab_{\ep\Qbar}, \quad \dab_{\epbar\cQ}, \quad
\dab_{\ep\Sbar}, \quad \dab_{\epbar\cS}, \quad
\dab_{\mc{P}^{-}}, \quad \dab_{\mc{J^{-}}}, \quad \dab_{\mc{K}^{-}}\ .
\eea
We find that (\ref{QQisP}) together with
\bea
\label{dyncon}
&&
[\dab_{\ep\Qbar},\dab_{\ep^\prime\Qbar}]\phi^a=0, \quad{}
[\dab_{\ep\Qbar},\dab_{\ep^\prime\Sbar}]\phi^a=0, \quad{}
[\dab_{\ep\Qbar},\dab_{\epbar\cS}]\phi^a=-i\epbar_m\ep^m\dab_{\mc{J^{-}}}\phi^a, \nn\\
&&
[\dab_{\mc{P}^{-}},\dab_{\ep\Qbar}]\phi^a=0, \quad{}
[\dab_{\mc{J}^{-}},\dab_{\ep\Qbar}]\phi^a=0
\eea
forms a set of independent dynamical constraints, with the rest of them following upon using the Jacobi identities. The last constraint, $[\dab_{\mc{J}^{-}},\dab_{\ep\Qbar}]\phi^a=0$, can equivalently be replaced by
\bea
\label{PScon}
[\dab_{\mc{P}^{-}},\dab_{\ep\Sbar}]\phi^a=0.
\eea
The dynamical bosonic constraint 
\bea
\label{PJcon}
[\dab_{\mc{P}^{-}},\dab_{\mc{J}^{-}}]\phi^a=0\ ,
\eea
follows from (\ref{dyncon}); it plays a central role, since all other bosonic dynamical constraints,
\bea
[\dab_{\mc{P}^{-}},\dab_{\mc{K}^{-}}]\phi^a=0, \quad{}
[\dab_{\mc{J}^{-}},\dab_{\mc{K}^{-}}]\phi^a=0\ ,
\eea
are derived from it by commuting with $\da_K$ and using JAC$(\da_K,\dab_{\mc{P}^{-}},\dab_{\mc{J}^{-}})$ and JAC$(\da_K,\dab_{\mc{P}^{-}},\dab_{\mc{K}^{-}})$.
We will use it to further restrict the form of the supersymmetry transformations. 

\subsection{Superspace BLG Theory}
In the $d=4$, $N=4$ SuperYang-Mills case, the dynamical supersymmetry transformations were fixed \emph{uniquely}~\cite{ANANTH} by solving the constraint (\ref{PJcon}). In the case at hand, this constraint will give us the BLG solution, although not quite uniquely.

The calculation of $\dab_{\mc{J}^{-}}\phi^a$ and then $[\dab_{\mc{P}^{-}},\dab_{\mc{J}^{-}}]\phi^a$ is quite lengthy. (See Appendices~\ref{App:P-J-} and \ref{App:[P-,J-]}.) Here we simply state the result for the odd case
\bea
\label{PJodd}
[\dab_{\mc{P}^{-}}^{\rm odd},\dab_{\mc{J}^{-}}^{\rm odd}]\phi^a=-\qter\mc{S}
\Big(\mc{F}\mc{O}_1^a+\mc{G}\mc{O}_2^a\Big)
+O(f^2)
\eea
where $\mc{S}$, $\mc{F}$ and $\mc{G}$ are $\p^{+}$-insertion operators, with $\mc{S}$ given in (\ref{defST}) and
\bea
\label{defFG}
\mc{F}=B^\star\wh{\mc{U}}_1+M^\star\wh{\mc{U}}_2, \quad
\mc{G}=C^\star\wh{\mc{U}}_3-D^\star\wh{\mc{U}}_4\ .
\eea
The coefficients
\bea
\label{star}
&&
B^\star \equiv B_\al+\al-\frac{5}{2}, \quad
M^\star \equiv M_\al-C_\al-D_\al+3 ,\nn\\
&&
C^\star \equiv C_\al-\al-2, \quad
D^\star \equiv D_\al-\al-2
\eea
are $\al$-independent thanks to the recursion relation (\ref{rec2}). This allowed us to pull them outside the sums in~\footnote{The sums in (\ref{O12}) contain $\cK_\al^a$ with $\al=3/2$ and $5/2$, which are outside the range of $\al$ for which $\cK_\al^a$ were originally introduced in (\ref{evenan}) and (\ref{oddan}). These $\cK_\al^a$ are \emph{defined} by the recursion relation (\ref{rec1}).
}
\bea
\label{O12}
\mc{O}_1^a &\equiv& \frac{\p^2}{\p r\p r^\prime}\Big\{ +\osum\left(
K_\al^{a\,[r+r^\prime,1]}-K_{\al+1}^{a\,[1, r+r^\prime]}\right)
+2\esum K_{\al+\half}^{a\,[r, r^\prime]} \Big\}\Big|_{\eta=\zeta=r=r^\prime=0}\nn\\[2pt]
\mc{O}_2^a &\equiv& \frac{\p^2}{\p r\p r^\prime}\Big\{ -\esum\left(
K_{\al+\half}^{a\,[r+r^\prime,1]}-K_{\al+\frac{3}{2}}^{a\,[1, r+r^\prime]}\right)
+2\osum K_{\al+1}^{a\,[r, r^\prime]} \Big\}\Big|_{\eta=\zeta=r=r^\prime=0}\ , \qquad
\eea
where the transverse derivatives $\p$ appear via the pairwise insertions of $E_r$ and $E_{r^\prime}$. After performing the differentiations with respect to the parameters $r$, $r^\prime$, $\eta$ and $\zeta$, and setting them to zero, we find that (\ref{PJodd}) is a sum of terms with four $\dbar$'s and two $\p$'s distributed in all possible ways among the three superfields.

The corresponding result in the even case is obtained under the substitution (\ref{oddtoeven}).

We found two ways in which the commutator (\ref{PJodd}) can vanish.
\begin{itemize}
\item
The first one is manifest: choose the values of the exponents so that $\mc{F}=\mc{G}=0$, that is
\bea
\label{starzero}
B^\star=M^\star=C^\star=D^\star=0\ .
\eea
Noting the dimensional constraint (\ref{ABMCD}), this corresponds to
\bea
A_{-\half}=2, \quad B_{-\half}=3, \quad M_{-\half}=0, \quad C_{-\half}=D_{-\half}=\frac{3}{2} \ .
\eea
As $C_\al=D_\al$, this imposes $[c d]$ antisymmetry on the structure constants, $f^a{}_{b c d}=-f^a{}_{b d c}$, since the symmetric part drops out in (\ref{oddan}). In the even case, (\ref{starzero}) also corresponds to a solution with
\bea \label{eventrivial}
A_{-1}=\frac{5}{2}, \quad B_{-1}=\frac{7}{2}, \quad M_{-1}=-1, \quad C_{-1}=D_{-1}=1.
\eea
This time $C_\al=D_\al$ implies $f^a{}_{b c d}=+f^a{}_{b d c}$, as the antisymmetric part drops out in (\ref{evenan}). Notice, however, that these solutions have \emph{fractional} powers of $\p^{+}$!
The fractional solutions have been reported earlier by one of us \cite{PRtalk}. We do not know if it is possible to make sense of such solutions. If they survive all the dynamical constraints (\ref{dyncon}) at $O(f)$, which we have not verified, one would have to go to $O(f^2)$ and see if it is still possible to satisfy all the constraints. Even if these solutions lead to algebraically consistent theories, their covariant formulations would likely contain square roots of invariant operators, such as $\sqrt {\p_\mu\p^\mu}$, and lead to non-locality. In this paper, we do not consider this type of solution any further.
%
\item
If we allow only \emph{integer} values of the exponents, then we find (see Appendix~\ref{App:odd}) that the \emph{only} way to make (\ref{PJodd}) vanish is to choose
\bea
\label{starBLG}
B^\star=0, \quad M^\star=-1, \quad C^\star=D^\star=-\half \ ,
\eea
corresponding to
\bea
\label{BLGexp}
A_{-\half}=B_{-\half}=3, \quad M_{-\half}=-2, \quad C_{-\half}=D_{-\half}=1 \ ,
\eea
\emph{and} require total antisymmetry of $f^a{}_{b c d}$ under the interchange of the last \emph{three} indices,
\bea
\label{ftasym}
f^a{}_{b c d}=f^a{}_{[b c d]}\ .
\eea
We found this by considering a particular subset of terms in (\ref{PJodd}) with all four $\dbar$'s and both $\p$'s acting on the same superfield. Under the above conditions, and \emph{only} in this case, the net contribution of these terms vanishes. The vanishing of the other subsets follow from the kinematical supersymmetry and the other linear symmetry transformations. We find that (\ref{BLGexp}) together with (\ref{ftasym}) correspond to the covariantly formulated BLG theory, which is known to be algebraically consistent. 
\end{itemize}
Indeed, with the values of the exponents in (\ref{BLGexp}) and using the antisymmetry property (\ref{ftasym}), we find (see Appendix~\ref{App:BLGform}) that (\ref{ansatzbar}) reduces to~\footnote{
In the $d=4$ SuperYang-Mills case, even after choosing the light-cone gauge, there remains a residual gauge symmetry on the transverse vector fields with gauge parameter satisfying $\p^{+}\Lambda=0$~\cite{DB}. As a result, the interacting supersymmetry transformations are obtained by covariantizing the transverse derivative~\cite{ANANTH}. In the $d=3$ SuperChern-Simons case (the BLG theory), there is no such residual symmetry (as the transverse vector fields are not independent degrees of freedom in the light-cone gauge~\cite{BENGT}), and we are unable to write the interacting transformations by generalizing the transverse derivative in (\ref{freeds}).
}
\bea
\label{BLGsusy}
\dab_{\epbar\cQ}^{int}\phi^a=-8\,\epbar_m\,f^a{}_{b c d}\,\frac{1}{\p^{+}}\left(
\p^{+}\phi^b\cdot\frac{1}{\p^{+}}\left((
\sqrt2\p^{+}\phi^c)(\p^{+}d^m\phibar^d)-i(\p^{+}\dbar_n\phi^c)( d^{mn }\phibar^d)\right)\right)\ ,
\eea
where four $\dbar$'s are absorbed into the conjugated superfield $\phibar^d$. Note that the two terms are required by chirality. 
After a rescaling of $f^a{}_{b c d}$, this matches the corresponding expression in \cite{DB} derived by direct light-cone gauge fixing of the BLG theory. The expression for $\dab_{\ep\Qbar}^{int}\phi^a$ following from (\ref{oddan}) is much more complicated. However, the inverse of (\ref{IOC}),
\bea
\label{barsusy}
\dab_{\ep\Qbar}^{int}\phi^a=\frac{\dbar{}^{[4]}}{2\p^{+2}}\left(\da_{\epbar\cQ}^{int}\phi^a\right)^\ast\ ,
\eea
provides an alternative (and compact) expression for it. 

As we show in Appendix \ref{App:even}, there is no such integer solution in the even case.

To summarize, we found that the odd Ansatz (\ref{oddan}) yields the BLG theory for the interaction part of the dynamical supersymmetry with the values of the exponents given in (\ref{BLGexp}) and the coefficients $f^a{}_{b c d}$ satisfying the antisymmetry condition (\ref{ftasym}). 

The basic result of this paper is that this is \emph{the only} solution to the constraints of the $OSp(2,2|8)$ superalgebra if we allow only integer powers of $\p^{+}$. 

Having matched the solution (\ref{BLGexp}) with the BLG theory, we have found that all the dynamical constraints (\ref{dyncon}) are satisfied at $O(f)$ thanks to the antisymmetry of $f^a{}_{b c d}$, and at $O(f^2)$ thanks to the Fundamental Identity \cite{BAG, GUS, Gran:2008vi} 
\bea
\label{FI}
f^a{}_{b c [d}f^b{}_{e f g]}=0,
\eea
which identifies $f^a{}_{b c d}$ with the structure constants of a 3-Lie algebra. Note that this symmetry is a global one in our formalism. There is no gauge field in the algebra. In the light-cone formulation this follows since the gauge field can be completely integrated out after the gauge fixing~\cite{BENGT}. Note also that at this level we have not obtained any quantization constraint on the structure constant. We expect that to happen when we further analyse the quantum properties of the theory.

The knowledge of the dynamical supersymmetry transformations fixes the theory uniquely, with all other dynamical transformations following by commutations. In particular, calculating the Hamiltonian shift $\dab_{\mc{P}^{-}}\phi^a$ using (\ref{QQisP}) determines the full interacting equations of motion,
\bea
\p^{-}\phi^a=i \dab_{\mc{P}^{-}}\phi^a\ .
\eea

\setcounter{equation}{0}
\section{BLG Hamiltonian as a Quadratic Form}\label{sec:BLG}
The full dynamical supersymmetry transformations in the light-cone superspace formulation of the BLG theory are given by the sum of the free transformations (\ref{freeds}),
\bea
\dab_{\epsbar\cQ}^{free}\phi^a=\frac{1}{\sqrt2}\epsbar_m q^m\frac{\p}{\p^{+}}\phi^a, \quad
\dab_{\eps\Qbar}^{int}\phi^a=\frac{1}{\sqrt2}\eps^m\qbar_m\frac{\p}{\p^{+}}\phi^a
\eea
and the interaction parts (\ref{BLGsusy}) and (\ref{barsusy}). Using (\ref{QQisP}), we can now find the complete BLG Hamiltonian shift $\dab_{\mc{P}^{-}}\phi^a$. Its free part is given by the standard expression (\ref{dyngen})
\bea
\dab_{\mc{P}^{-}}^{free}\phi^a=-\frac{i}{2}\frac{\p^2}{\p^{+}}\phi^a.
\eea
To write the corresponding light-cone superspace Hamiltonian $H$, we need to introduce a metric $h_{a b}=h_{b a}$ for the gauge indices. Then the free Hamiltonian is
\bea
\label{freeH}
H^{free}=h_{a b}\int dz\phibar^a\frac{\p^2}{\p^{+2}}\phi^b,
\eea
where $dz=d^3x d^4\theta d^4\tabar$, and it is related to the Hamiltonian shift via the functional derivative,
\bea
\label{HfromP}
\frac{\da H^{free}}{\da\phi^a}=8i h_{a b}\p^{+}(\dab_{\mc{P}^{-}}^{free}\phi^b),
\eea
as can be easily verified using the basic rule of functional differentiation \cite{Bengtsson:1983pg}
\bea
\frac{\da\phi^a(z)}{\da\phi^b(z^\prime)}=d^{[4]}\da(z-z^\prime)\da^a_b
\eea
and the inside-out constraint. The full Hamiltonian $H$ can then be found by integrating (\ref{HfromP}) with the full Hamiltonian shift $\dab_{\mc{P}^{-}}\phi^a$. However, instead of doing the complicated integration, we can start with a natural guess for $H$ and verify that it yields the correct $\dab_{\mc{P}^{-}}\phi^a$ upon differentiation. Such a guess is provided by the quadratic form property of the light-cone superspace Hamiltonian in maximally supersymmetric theories, discovered in \cite{ANANTH}. If this property holds in the BLG theory, then we should have
\bea
\label{HisQF}
H &=& \frac{i}{\sqrt2}h_{a b}\int dz\Qbar_m^a\frac{1}{\p^{+}}\cQ^{b m} 
\equiv \frac{1}{\sqrt2}\left<\cQ,\cQ\right> \nn\\
&=& \frac{i}{2\sqrt2}h_{a b}\int dz\left(
\qbar_m\frac{\p}{\p^{+}}\phibar^a+\Wbar_m^a\right)\frac{1}{\p^{+}}\left(
q^m\frac{\p}{\p^{+}}\phi^b+\cW^{b m}\right),
\eea
where the hermitian form $\left<\ ,\ \right>$ is defined in (\ref{hqf}). The $\cQ^{a m}$ and $\cW^{a m}$ are defined by
\bea
\dab_{\epbar\cQ}\phi^a \equiv \epbar_m\cQ^{a m}, \quad
\dab_{\epbar\cQ}^{int}\phi^a \equiv \frac{1}{\sqrt2}\epbar_m\cW^{a m}
\eea
together with $\Qbar_m^a \equiv (\cQ^{a m})^\ast$ and $\Wbar_m^a \equiv (\cW^{a m})^\ast$. From (\ref{BLGsusy}), we have, explicitly,
\bea
\cW^{a m} &=& 8f^a{}_{b c d}\frac{1}{\p^{+}}\left(\p^{+}\phi^b\cdot\frac{1}{\p^{+}}\left(
2\p^{+}d^m\phibar^c\cdot\p^{+}\phi^d-i\sqrt2\, d^{m n}\phibar^c\cdot\p^{+}\dbar_n\phi^d\right)\right), \nn\\
\Wbar^a_m &=& 8f^a{}_{b c d}\frac{1}{\p^{+}}\left(\p^{+}\phibar^b\cdot\frac{1}{\p^{+}}\left(
2\p^{+}\dbar_m\phi^c\cdot\p^{+}\phibar^d-i\sqrt2\,\dbar_{m n}\phi^c\cdot\p^{+}d^n\phibar^d\right)\right).
\eea
At the free level, after several integrations by parts, the use of the inside-out constraint (\ref{IOC}) and the anticommutator (\ref{qqbar}), we find that (\ref{HisQF}) reproduces $H^{free}$ in (\ref{freeH}). For the $O(f)$ part of $H$, we have
\bea
\label{H1}
H^{(1)}=\frac{i}{2\sqrt2}h_{a b}\int dz\left(
q^m\frac{\p}{\p^{+}}\phibar^a\cdot\frac{1}{\p^{+}}\Wbar^b_m
\right)+c.c.
\eea
whereas the $O(f^2)$ part is 
\bea
\label{H2}
H^{(2)}=\frac{i}{2\sqrt2}h_{a b}\int dz\left(
\Wbar_m^a\frac{1}{\p^{+}}\cW^{b m} \right)
\eea
We have \emph{not} verified that functional differentiation of $H^{(1)}$ and $H^{(2)}$ reproduces $\dab_{\mc{P}^{-}}^{(1)}\phi^a$ and $\dab_{\mc{P}^{-}}^{(2)}\phi^a$ as follow from (\ref{QQisP}). By analogy with the $d=4$ SuperYang-Mills \cite{ANANTH}, we expect that such a verification at $O(f)$ would require \emph{total} antisymmetry of the structure constants,
\bea
\label{ft4asym}
f_{a b c d} \equiv h_{a g}f^g{}_{b c d}=f_{[a b c d]}
\eea
whereas at $O(f^2)$ the Fundamental Identity (\ref{FI}) would be required. What we \emph{have} verified, is that when $H^{(1)}$ is written in terms of the component fields $A^a$, $C^{m n a}$, $\chi^{m a}$ (and their conjugates), its ``$C$-only'' part, after using (\ref{ft4asym}), is
\bea
\label{H1C}
H^{(1)}{}_{\big|\text{$C$-only}} =-16 f_{a b c d}\int d^3x
(\Cbar{}^a_{i j}\p C^{i j b})\frac{1}{\p^{+}}
(\Cbar{}^c_{m n}\p^{+}C^{m n d})
\eea
which matches the corresponding part in the light-cone BLG Hamiltonian \cite{BENGT}. 
This is enough to show that (\ref{H1}) can be transformed to the form proposed by Nilsson \cite{BENGT},
\bea
\label{H1BN}
H^{(1)} =-8 f_{a b c d}\int dz(\phi^a\p\phi^b)\frac{1}{\p^{+}}(\phibar^c\p^{+}\phibar^d)+c.c.
\eea
as the two expressions match on the level of ``$C$-only'' terms. (This way of verifying equivalence of two different superfield expressions was also used in \cite{ANANTH}.) As the calculations involved are quite nontrivial (see Appendix~\ref{App:HisQF}), we feel that this provides sufficient evidence for the correctness of the full Hamiltonian given as the quadratic form (\ref{HisQF}).

\section{Conclusion and Discussion}
In this paper, we have constructed the superconformal theory of Bagger, Lambert and Gustavsson in three dimensions by requiring closure of $OSp(2,2|8)$ on constrained chiral superfields in light-cone superspace. The algebra splits into kinematical and dynamical operators: kinematical operators act linearly on the superfields, while dynamical operators contain terms linear (free theory) and non-linear (interactions) in the superfields. 

A feature of any superconformal theory is that all dynamics is algebraically determined by its supersymmetry transformations. We first determined ans\"atze for the dynamical supersymmetry transformations which satisfied all kinematical constraints. These constraints \emph{required} that the theory contain a fourth order tensor $f^{\,a}_{~\,bcd}$.  By demanding commutation of the transformations generated by the Hamiltonian and boost, we were able to narrow down the form of the supersymmetry transformations to two choices. 
\vskip .3cm
One is the BLG theory, which requires the antisymmetry of the $f^{\,a}_{~\,[bcd]}$ type.
\vskip .3cm
The other ``solutions'' to the algebraic constraints entail fractional powers of light-cone derivatives $\p^{+}$, and partial symmetries whereby  $f^{\,a}_{~\,bcd}$ are symmetric (antisymmetric) under $c\leftrightarrow d$ for the even (odd) cases. In this paper, we have not checked their consistency with the full algebra, since their covariant formulation would likely lead to square roots of covariant operators such as $\sqrt{\p_\mu\p^\mu}$, and therefore non-local interactions. 

\vskip .3cm
Our formulation of the BLG theory has many analogies to $N=4 $ SuperYang-Mills, since they use the same  chiral superfield. In particular, the   light-cone superspace Hamiltonian of both theories can be written as a quadratic form. 
In the Yang-Mills case we found a tensor of the form $f^{\,a}_{~\,bc}$ which satisfied Lie algebra Jacobi identities from closure of the algebra. In the BLG case, we found  $f^{\,a}_{~\,bcd}$, which will satisfy the fundamental identity of BLG, from closure as well\footnote{In a similar approach,  the authors of  \cite{Samtleben:2009ts} considered standard $N=8$ superspace constraints and also obtained the BLG theory as a special solution described by the same fourth rank tensor $f^{a}{}_{[bcd]}$.
}. 
Our formalism has invoked $SO(8)$ as the $R$-symmetry. We can now use this formalism and relax part of the $R$-symmetry to search for other superconformal theories.

The same chiral superfield in $d=5$ and $d=6$ dimensions forms a linear representation of the superconformal group appropriate to these dimensions. We intend to use these algebraic techniques to study their possible interactions in future publications.

\vskip 1cm
\noindent
{\bf\large Acknowledgements}

\noindent
We thank Renata Kallosh and Bengt Nilsson for useful discussions and comments. SK thanks Sungjay Lee, Jian Qiu, and Piljin Yi for useful discussions. 
LB would like to thank the Institute for Fundamental Theory at the University of Florida 
for hospitality during his visit. SK also thanks Sungkyunkwan Univerisity, Korean Institute 
for Advanced Study, and the Institute  for Fundamental Theory at the University of Florida for hospitality during his visit. DB thanks Jonathan
Bagger and Warren Siegel for helpful discussions.
The research of DB and PR is partially supported by the Department of Energy Grant No. DE-FG02-97ER41029. 
SK is supported by the Universit\'{e} Libre de Bruxelles and the
International Solvay Institutes, and by IISN - Belgium convention
4.4505.86, and by the Belgian Federal Science Policy Office through
the Interuniversity Attraction Pole P5/27.


\section*{Appendix}
\appendix
\makeatletter

\setcounter{equation}{0}
\section{Light-Cone $OSp(2,2|8)$ Algebra }\label{App:algebra}

The free theory (operator) representation of the $OSp(2,2|8)$ superalgebra used in this paper is given by
\bea
&& P^{+}=-i\p^{+}, \quad P=-i\p, \quad 
\mc{P}^{-}=-\frac{i}{2}\frac{\p^2}{\p^{+}}, \nn\\
&& J^{+}=i x\p^{+}, \quad J^{+-}=i(\mc{A}+\frac{x}{2}\p+\half), \quad 
\mc{J}^{-}=-i\frac{\p}{\p^{+}}\mc{A}, \nn\\
&& D=i(\mc{A}-\frac{x}{2}\p), \quad K^{+}=i x^2\p^{+}, \quad K=2i x\mc{A}, \quad
\mc{K}^{-}=2i\frac{1}{\p^{+}}\mc{A}(\mc{A}-\half), \nn\\
&& T^m{}_n=\frac{i}{\sqrt2\p^{+}}\left(q^m\qbar_n-\qter\da^m{}_n q^k\qbar_k\right), \quad
T=\frac{i}{4\sqrt2\p^{+}}(q^k\qbar_k-\qbar_k q^k), \nn\\
&& T^{m n}=\frac{i}{\sqrt2\p^{+}}q^m q^n, \quad
\Tbar_{m n}=\frac{i}{\sqrt2\p^{+}}\qbar_m\qbar_n, \nn\\
&&  \cQ^m=\frac{1}{\sqrt2}\frac{\p}{\p^{+}}q^m, \quad
s^m=i x q^m, \quad \cS^m=-i q^m\frac{1}{\p^{+}}\mc{A}, \nn\\
&&  \Qbar_m=\frac{1}{\sqrt2}\frac{\p}{\p^{+}}\qbar_m, \quad
\sbar_m=-i x\qbar_m, \quad \Sbar_m=i\qbar_m\frac{1}{\p^{+}}\mc{A}
\eea
together with the kinematical supersymmetry generators,
\bea
&& q^m=-\frac{\p}{\p\tabar_m}+\frac{i}{\sqrt2}\ta^m\p^{+}, \quad
\qbar_m=\frac{\p}{\p\ta^m}-\frac{i}{\sqrt2}\tabar_m\p^{+}, \nn\\
&& \mc{A}=x^{-}\p^{+}-\frac{x}{2}\p-\half\mc{N}+\half, \quad
\mc{N}=\ta^k\frac{\p}{\p\ta^k}+\tabar_k\frac{\p}{\p\tabar_k}.
\eea
The generators of the conformal group are chosen to be \emph{hermitian} with respect to the following hermitian form~\footnote{
Complex conjugation, denoted by $\ast$, interchanges the order of operands: $(\mc{O}_1\dots\mc{O}_n)^\ast=\mc{O}_n^\ast\dots\mc{O}_1^\ast$ irrespective of whether $\mc{O}$'s are bosonic or fermionic objects. See Appendix A in [4] for more details. However, our rule for conjugation of \emph{fermionic parameters} differs from \cite{DB}: we define $(\eps^m)^\ast=-\epsbar_m$ so that $(\eps^m\qbar_m\phi)^\ast=+\epsbar_m q^m\phibar$.
}
\bea
\label{hqf}
\left<\phi_1,\phi_2\right>=i \int d^3x d^4\ta d^4\tabar \phi_1^\ast\frac{1}{\p^{+}}\phi_2.
\eea
The hermitian conjugate $\mc{O}^\dagger$ of an operator $\mc{O}$ is defined by
\bea
\left<\phi_1,\mc{O}\phi_2\right>=\left<\mc{O}^\dagger\phi_1,\phi_2\right>
\eea
so that $\mc{O}^\dagger=\mc{O}^\ast$ if $\mc{O}$ does not depend on $x^{-}$. The dependence on $x^{-}$ in the above generators comes via the dependence on $\mc{A}$. Direct computation gives
\bea
\mc{A}^\dagger=-\mc{A}-\half
\eea
and then hermiticity properties of all the generators follow,
\bea
\label{hcgen}
\ba[b]{rclcl}
\mc{O}^\dagger &=& +\mc{O} &&\text{for }\mc{O}=(P^{+},P,\mc{P}^{-},J^{+},J^{+-},\mc{J}^{-},D,K^{+},K,\mc{K}^{-}) \\
(\mc{O}^m)^\dagger &=& -\ov{\mc{O}}_m &&\text{for }\mc{O}^m=(q^m,\cQ^m,s^m,\cS^m) 
\ea\nn\\
(T^m{}_n)^\dagger=T^n{}_m, \quad (T)^\dagger=T, \quad (T^{m n})^\dagger=-\Tbar_{m n}. \hspace{120pt}
\eea
Using the following basic commutation properties
\bea
\label{Acomm}
&& [\p^{+},x^{-}]=-1, \quad{} [\p,x]=1, \quad \{q^m,q^n\}=\{\qbar_m,\qbar_n\}=0, \nn\\[5pt]
&& \{q^m,\qbar_n\}=i\sqrt2\da^m{}_n\p^{+}, \quad{}
[\mc{A},q^m]=\half q^m, \quad{} [\mc{A},\qbar_m]=\half\qbar_m, \nn\\ [5pt]
&& [\mc{A},\p^{+}]=\p^{+}, \quad{} [\mc{A},\frac{1}{\p^{+}}]=-\frac{1}{\p^{+}}, \quad{}
[\mc{A},x]=-\half x, \quad{} [\mc{A},\p]=\half\p 
\eea
one can verify that the algebra closes. The \emph{non-vanishing} (anti)commutators of $OSp(2,2|8)$ are as follows.~\footnote{We note that $P$'s commute with $P$'s, $K$'s commute with $K$'s, $D$ commutes with $J$'s, and $R$-symmetry generators $T$'s commute with all other bosonic generators.} 
\begin{itemize}
\item
In the $Sp(2,2)\sim SO(3,2)$ conformal group sector:
\bea
&&
[J^{+-},J^{+}]=i J^{+}, \quad{}
[J^{+-},\mc{J}^{-}]=-i\mc{J}^{-}, \quad{}
[J^{+},\mc{J}^{-}]=i J^{+-}. \nn\\[5pt]
&&
[J^{+-},P^{+}]=i P^{+}, \quad{}
[J^{+-},\mc{P}^{-}]=-i\mc{P}^{-}, \quad{}
[J^{+-},K^{+}]=i K^{+}, \quad{}
[J^{+-},\mc{K}^{-}]=-i\mc{K}^{-}. \hspace{10pt} \nn\\[5pt]
&&
[J^{+},P]=-i P^{+}, \quad{}
[J^{+},\mc{P}^{-}]=-i P, \quad{}
[J^{+},K]=-i K^{+}, \quad{}
[J^{+},\mc{K}^{-}]=-i K. \nn\\[5pt]
&&
[J^{-},P^{+}]=-i P, \quad{}
[J^{-},P]=-i\mc{P}^{-}, \quad{}
[J^{-},K^{+}]=-i K, \quad{}
[J^{-},K]=-i K^{-}. \nn\\[5pt]
&&
[K^{+},P]=2i J^{+}, \quad{}
[K^{+},\mc{P}^{-}]=2i(J^{+-}-D). \nn\\[5pt]
&&
[\mc{K}^{-},P]=2i\mc{J}^{-}, \quad{}
[\mc{K}^{-},P^{+}]=-2i(J^{+-}+D). \nn\\[5pt]
&&
[K,P^{+}]=-2i J^{+}, \quad{}
[K,P^{-}]=-2i J^{-}, \quad{}
[K,P]=2i D. \nn\\[5pt]
&&
[D,P^{+}]=i P^{+}, \quad{} 
[D,P]=i P, \quad{}
[D,\mc{P}^{-}]=i\mc{P}^{-}. \nn\\[5pt]
&&
[D,K^{+}]=-i K^{+}, \quad{}
[D,K]=-i K, \quad{}
[D,\mc{K}^{-}]=-i\mc{K}^{-}.
\eea
\item
In the $SO(8)$ $R$-symmetry group sector:
\bea
&& 
[T^m{}_n,T^k{}_l]=\da^m{}_l T^k{}_n-\da^k{}_n T^m{}_l ,\nn\\[5pt]
&& 
[T^m{}_n,T^{k l}]=\half\da^m{}_n T^{k l}-\da^k{}_n T^{m l}+\da^l{}_n T^{m k}, \quad{}
[T,T^{m n}]=-T^{m n}. \nn\\[2pt]
&&
[T^n{}_m,\Tbar_{k l}]=\half\da^n{}_m \Tbar_{k l}-\da^n{}_k \Tbar_{m l}+\da^n{}_l \Tbar_{m k}, \quad{}
[T,\Tbar_{m n}]=-\Tbar_{m n}. \nn\\[5pt]
&& [T^{m n},\Tbar_{k l}]=\da^m{}_k T^n{}_l-\da^m{}_l T^n{}_k+\da^n{}_l T^m{}_k-\da^n{}_k T^m{}_l 
+(\da^m{}_k\da^n{}_l-\da^m{}_l\da^n{}_k)T. \hspace{50pt}
\eea
\item
$R$-symmetry group action on the fermionic generators:
\bea
&&
[T^m{}_n,q^k]=\qter\da^m{}_n q^k-\da^k{}_n q^m, \quad{}
[T,q^m]=-\half q^m, \quad{}
[\Tbar_{m n},q^k]=\da^k{}_m\qbar_n-\da^k{}_n\qbar_m, \hspace{20pt} \nn\\
&&
[T^n{}_m,\qbar_k]=-\qter\da^n{}_m\qbar_k+\da^n{}_k\qbar_m, \quad{}
[T,\qbar_m]=\half\qbar_m, \quad{}
[T^{m n},\qbar_k]=\da^m{}_k q^n-\da^n{}_k q^m \nn\\[5pt]
&&
\eea
and identically for $s^m$ and $\sbar_m$, $\cQ^m$ and $\Qbar_m$, $\cS^m$ and $\Sbar_m$.
\item
Conformal group action on the fermionic generators:
\bea
\ba[b]{rclcrcl}
[J^{+-},q^m] &=&\dst \frac{i}{2}q^m 
&\hspace{40pt}& 
[J^{+-},\qbar_m] &=&\dst \frac{i}{2}\qbar_m 
\\[5pt]
[J^{+-},\cQ^m] &=&\dst -\frac{i}{2}\cQ^m
&& 
[J^{+-},\Qbar_m] &=&\dst -\frac{i}{2}\Qbar_m 
\\[5pt]
[J^{+-},s^m] &=&\dst \frac{i}{2}s^m 
&&
[J^{+-},\sbar_m] &=&\dst \frac{i}{2}\sbar_m 
\\[5pt]
[J^{+-},\cS^m] &=&\dst -\frac{i}{2}\cS^m 
&&
[J^{+-},\Sbar_m] &=&\dst -\frac{i}{2}\Sbar_m
\\[5pt]
[J^{+},\cQ^m] &=&\dst -\frac{i}{\sqrt2}q^m
&&
[J^{+},\Qbar_m] &=&\dst -\frac{i}{\sqrt2}\qbar_m
\\[5pt]
[J^{+},\cS^m] &=&\dst \frac{i}{2}s^m
&&
[J^{+},\Sbar_m] &=&\dst \frac{i}{2}\sbar_m
\\[5pt]
[\mc{J}^{-},q^m] &=&\dst -\frac{i}{\sqrt2}\cQ^m
&&
[\mc{J}^{-},\qbar_m] &=&\dst -\frac{i}{\sqrt2}\Qbar_m
\\[5pt]
[\mc{J}^{-},s^m] &=& i\cS^m
&&
[\mc{J}^{-},\sbar_m] &=& i\Sbar_m
\\[5pt]
[P^{+},\cS^m] &=& q^m 
&& 
[P^{+},\Sbar_m] &=& -\qbar_m 
\\[5pt]
[P,s^m] &=& q^m 
&& 
[P,\sbar_m] &=& -\qbar_m 
\\[5pt]
[P,\cS^m] &=&\dst \frac{1}{\sqrt2}\cQ^m 
&& 
[P,\Sbar_m] &=&\dst -\frac{1}{\sqrt2}\Qbar_m 
\\[10pt]
[\mc{P}^{-},s^m] &=& \sqrt2\cQ^m 
&& 
[\mc{P}^{-},\sbar_m] &=& -\sqrt2\,\Qbar_m 
\\[5pt]
[K^{+},\cQ^m] &=& -\sqrt2 s^m
&&
[K^{+},\Qbar_m] &=& \sqrt2\sbar_m
\\[5pt]
[K,q^m] &=& s^m
&&
[K,\qbar_m] &=& -\sbar_m
\\[5pt]
[K,\cQ^m] &=& \sqrt2\cS^m
&&
[K,\Qbar_m] &=& -\sqrt2\,\Sbar_m
\\[5pt]
[\mc{K}^{-},q^m] &=& -2\cS^m
&&
[\mc{K}^{-},\qbar_m] &=& 2\Sbar_m
\\[5pt]
[D,q^m] &=&\dst \frac{i}{2}q^m
&&
[D,\qbar_m] &=&\dst \frac{i}{2}\qbar_m
\\[5pt]
[D,\cQ^m] &=&\dst \frac{i}{2}\cQ^m
&&
[D,\Qbar_m] &=&\dst \frac{i}{2}\Qbar_m
\\[5pt]
[D,s^m] &=&\dst -\frac{i}{2}s^m
&&
[D,\sbar_m] &=&\dst -\frac{i}{2}\sbar_m
\\[5pt]
[D,\cS^m] &=&\dst -\frac{i}{2}\cS^m
&&
[D,\Sbar_m] &=&\dst -\frac{i}{2}\Sbar_m.
\ea
\eea
\item
Anticommutation relations:
\bea
\ba[b]{rclcrcl}
\{q^m,\qbar_n\} &=& -\sqrt2\da^m{}_n P^{+} 
&&
\{\cQ^m,\Qbar_n\} &=& -\sqrt2\da^m{}_n\mc{P}^{-} 
\\[5pt]
\{s^m,\sbar_n\} &=& \sqrt2\da^m{}_n K^{+}
&&
\{\cS^m,\Sbar_n\} &=&\dst \frac{1}{\sqrt2}\da^m{}_n\mc{K}^{-}
\\[10pt]
\{q^m,\Qbar_n\} &=& -\da^m{}_n P
&&
\{\qbar_m,\cQ^n\} &=& -\da^n{}_m P
\\[5pt]
\{q^m,\sbar_n\} &=& -i\sqrt2\da^m{}_n J^{+}
&&
\{\qbar_m,s^n\} &=& i\sqrt2\da^n{}_m J^{+}
\\[5pt]
\{\cQ^m,\Sbar_n\} &=& -i\da^m{}_n\mc{J}^{-}
&&
\{\Qbar_m,\cS^n\} &=& i\da^n{}_m\mc{J^{-}}
\\[5pt]
\{q^m,\cS^n\} &=&\dst \frac{1}{\sqrt2}T^{m n}
&&
\{\qbar_m,\Sbar_n\} &=&\dst -\frac{1}{\sqrt2}T_{m n}
\\[10pt]
\{\cQ^m,s^n\} &=& T^{m n}
&&
\{\Qbar_m,\sbar_n\} &=& -\Tbar_{m n}.
\ea \nn
\eea
\vspace{-30pt}
\bea
\{q^m,\Sbar_n\} &=&\dst \frac{i}{\sqrt2}(J^{+-}+D)\da^m{}_n-\frac{1}{\sqrt2}(T^m{}_n+\half T\da^m{}_n)
\nn\\[5pt]
\{\qbar_m,\cS^n\} &=&\dst -\frac{i}{\sqrt2}(J^{+-}+D)\da^n{}_m-\frac{1}{\sqrt2}(T^n{}_m+\half T\da^n{}_m) \hspace{23pt}
\nn\\[5pt]
\{\cQ^m,\sbar_n\} &=& -i(J^{+-}-D)\da^m{}_n-(T^m{}_n+\half T\da^m{}_n)
\nn\\[5pt]
\{\Qbar_m,s^n\} &=& i(J^{+-}-D)\da^n{}_m-(T^n{}_m+\half T\da^n{}_m).
\eea
\end{itemize}
This set of commutation relations is invariant under hermitian conjugation (\ref{hcgen}). When using these \emph{operator} commutation relations to write the corresponding ones for \emph{transformations}, one has to note the following minus sign
\bea
\label{minus}
[\da_{\mc{O}_1},\da_{\mc{O}_2}]\phi^a=-[\mc{O}_1,\mc{O}_2]\phi^a,
\eea
where $\da_{\mc{O}}\phi^a=\mc{O}\phi^a$ are \emph{free} theory transformations. The resulting set of commutation relations is required to be satisfied by the \emph{interacting} theory transformations as well.

\setcounter{equation}{0}
\section{Useful Identities}\label{App:useful}
We present a set of useful formulae and identities:

\begin{itemize}

\item Commutators:

\bea\label{com1}
[E_\eps,\ta^m\qbar_n]=\eps^m\frac{\p}{\p\eps^n}E_\eps, \quad{}
[E_\eta,\ta^m q_n]=\eta^m\left(\frac{\p}{\p\eta^n}-i\sqrt2\,\tabar_n\right)E_\eta\ .
\eea

\bea\label{com2}
[\mc{A},\dbar_m]~=~\half\dbar_m\ ,\qquad [E_\eps E_\eta\p^{+k},\mc{A}]=\left(\half\eps\frac{\p}{\p\eps}+\half\eta\frac{\p}{\p\eta}-k\right)E_\eps E_\eta\p^{+k}\ .
\eea

\bea
[E_\eta,\ta^m]=\wh\eta^m E_\eta, \quad{}
[E_\eta,\ta^m\ta^n\p^{+}]=(\ta^m\eta^n-\ta^n\eta^m+\eta^m\wh\eta^n)E_\eta.
\eea

\item The Master Formula: 

Consider the commutator of a transformation linear in $\phi^a$, $\da_{\mc{O}}\phi^a=\mc{O}\phi^a$, with a transformation nonlinear in $\phi$'s,
\bea
\dab_{X}\phi^a \equiv f^a{}_{b c d}\left((X_1\phi^b) X_2((X_3\phi^c)( X_4\phi^d))\right)\ , 
\eea
where $X_i$ are operators. In terms of the insertion operators, their commutator can be written as the master formula

\bea
\label{master}
[\da_{\mc{O}},\dab_X]\phi^a 
=\left(\sum_{i=1}^4[X_i,\mc{O}]X_i^{-1}{\mc U}_i+\{\mc{O}\}_{12}+\{\mc{O}\}_{34}\right)\dab_{X}\phi^a\ ,
\eea
where 
\bea
\label{triplet}
\{\mc{O}\}_{12} \equiv \mc{O}\mc{U}_1+\mc{O}\mc{U}_2-\mc{O}
\ ,\qquad 
\{\mc{O}\}_{34} \equiv \mc{O}\mc{U}_3+\mc{O}\mc{U}_4-\mc{U}_2\mc{O}\ ,
\eea
account for the deviation from Leibnitz's rule. Indeed if $\mc D$ is a derivative operator, then $
\{\mc{D}\}_{12}=\{\mc{D}\}_{34}=0$. Given two \emph{derivative} operators $\mc{D}$ and $\mc{D}^\prime$, commuting with $\p^{+}$, we find that (even if $\mc{D}$ and $\mc{D}^\prime$ do not commute),

\bea
\label{D1D2}
\left\{\frac{\mc{D}\mc{D}^\prime}{\p^{+}}\right\}_{ij}~=~\frac{\p}{\p r}\frac{\p}{\p r^\prime}\frac{1}{\p^{+}}
\Big((\p^{+} \mc{E}_{r} \mc{E}_{r^\prime}\mc{U} _i)\,(\p^{+}\mc{E}_{-r}\mc{E}_{-r^\prime} \mc{U}_j)\Big)\Big|_{r=r^\prime=0}\ ,
\eea
for $(ij)=(12), (34)$, and where 
\bea
\mc{E}_r \equiv e^{r\wh{\mc{D}}}, \quad
\mc{E}_{r^\prime} \equiv e^{r^\prime\wh{\mc{D}}^\prime}.
\eea
A frequently used identity is
\bea
\label{doublet}
\wh{\mc{O}} \mc{U}_i-\wh{\mc{O}} \mc{U}_j~=~\frac{\p}{\p r}\Big((e^{r\wh{\mc{O}}}\mc{U}_i)\,(e^{-r\wh{\mc{O}}} \mc{U}_j)\Big)
\Big|_{r=0}.
\eea
Noting that $\wh{\qbar}_m=\wh{\dbar}_m-i\sqrt2\,\tabar_m$ and $\wh q^m=\wh d^m+i\sqrt2\ta^m$, we have
\bea
\label{qisd}
(e^{\eps\,\wh\qbar}\mc{U}_i)\,(e^{-\eps\,\wh\qbar}\mc{U}_j)~=~(e^{\eps\,\wh\dbar}\mc{U}_i)\,(e^{-\eps\,\wh\dbar}\mc{U}_j)
, \quad
(e^{\epsbar\,\wh q}\mc{U}_i)\,(e^{-\epsbar\,\wh q}\mc{U}_j)~=~(e^{\epsbar\,\wh d}\mc{U}_i)\,(e^{-\epsbar\,\wh d}\mc{U}_j).
\eea
Using then $\{d^m,\dbar_n\}=-i\sqrt2\da^m{}_n\p^{+}$, and $d^m\phi^a=0$, we find that~\footnote{
In these appendices, it is implicitly assumed that only the terms linear in $\ep$ and $\epbar$ are kept. In addition, whenever the sums are involved, setting $\eta=\zeta=0$ after differentiations is also assumed.
}

\bea\label{shuffle}
\cK_\al^{a\,(\eps,\eta,\zeta)}=\ep^m\frac{\p}{\p\eta^m}\cK_\al^{a\,(0,\eta,\zeta)}, \quad
\cK_\al^{a\,(\epsbar,\eta,\zeta)}
~=~i\sqrt2\,\epsbar_m\eta^m\mc{S}\cK_\al^{a\,(0,\eta,\zeta)} .
\eea

\item Selected Applications\\
The master formula (\ref{master}) and use of (\ref{com1}) yields,
\bea
[\da_{SU(4)},\dab_{\ep\Qbar}^{int}]\phi^a &=& -\omega^m{}_n\frac{1}{\sqrt2}\sum\left(
\eps^n\frac{\p}{\p\eps^m}+\eta^n\frac{\p}{\p\eta^m}+\zeta^n\frac{\p}{\p\zeta^m}\right)
\mc{K}_\alpha^{a\,(\ep,\eta,\zeta)} ,\nn\\[5pt]
[\da_{U(1)},\dab_{\eps\Qbar}^{int}]\phi^a &=& -\omega \frac{1}{\sqrt2}\sum
\left(
\eps\frac{\p}{\p\eps}+\eta\frac{\p}{\p\eta}+\zeta\frac{\p}{\p\zeta}\right)
\mc{K}_\alpha^{a\,(\ep,\eta,\zeta)}
\ .
\eea
 
Since $\mc{A}$ is not a derivative operator, the master formula (\ref{master}) gets a contribution from the ``triplets'' (\ref{triplet}), and using (\ref{com2}), we obtain

\bea
[\da_{\mc{A}},\dab_{\ep\Qbar}^{int}]\phi^a &=& \frac{1}{\sqrt2}\sum\Big[
1+\half\left(\ep\frac{\p}{\p\ep}+\eta\frac{\p}{\p\eta}+\zeta\frac{\p}{\p\zeta}\right) \nn\\
&&\hspace{50pt}
+(A_\al-B_\al+M_\al-C_\al-D_\al)\Big]\cK_\al^{a\,(\eps,\eta,\zeta)}\ ,
\eea
which leads to the dimensional constraint (\ref{ABMCD}).

We also get 
\be
[\da_{coset},\dab_{\eps\Qbar}^{int}]\phi^a = -i\ombar_{m n}\eps^k\sum\frac{\p}{\p\eta^k}
\Big(\eta^m\eta^n(\wh{\mc{U}}_1+\wh{\mc{U}}_2)+\zeta^m\zeta^n(\wh{\mc{U}}_3+\wh{\mc{U}}_4)\Big)
\cK_\al^{a\,(0,\eta,\zeta)}\ , 
\ee
which after using (\ref{shuffle}) and moving the $\eta$-derivative, becomes
\be
 2i\ombar_{m n}\eps^n\sum\eta^m\mc{S}^{-1}\cK_\al^{a\,(0,\eta,\zeta)}
-i\ombar_{m n}\sum(\eta^m\eta^n\mc{S}^{-1}+\zeta^m\zeta^n\mc{T}^{-1})\cK_\al^{a\,(\eps,\eta,\zeta)}.
\ee
This yields (\ref{TQbar}).

\item  Even-Odd sum Relations\\
The $K_{\al}$'s defined in \eqref{compact}, satisfy identities which convert the even sum to the odd sum and vice versa 

\begin{align} 
\sum_{\rm  even}\frac{\d}{\d\eta^{m}}K_{\al}\, =\, - \sum_{\rm odd}\frac{\d}{\d\zeta^{m}}K_{\al+\frac12}\ ,\quad
\sum_{\rm  even}\eta^{m}K_{\al}\,=\,-\sum_{\rm odd}\zeta^{m}K_{\al-\frac12}\ ,\\
\sum_{\rm  odd}\frac{\d}{\d\eta^{m}}K_{\al}\, =\,  +\sum_{\rm  even}\frac{\d}{\d\zeta^{m}}K_{\al+\frac12}\ ,\quad
\sum_{\rm odd}\eta^{m}K_{\al}\,=\,+\sum_{\rm  even}\zeta^{m}K_{\al-\frac12}\ .
\end{align}

\begin{align}
\sum_{\rm  even} \left( \zeta^{m}\frac{\d}{\d\eta^{n}} K_{\al} - \eta^{m}\frac{\d}{\d\zeta^{n}} K_{\al+1}\right) &= \delta^{m}_{n}\sum_{\rm  odd} K_{\al+\frac12},\label{even-id}\\
\sum_{\rm  odd} \left( \zeta^{m}\frac{\d}{\d\eta^{n}}  K_{\al} - \eta^{m}\frac{\d}{\d\zeta^{n}} K_{\al+1}\right) &= -\delta^{m}_{n}\sum_{ \rm even}K_{\al+\frac12}\ .\label{odd-id}
\end{align}

\item Identities for alternate nesting of the supersymmetry parameters  

\begin{align}
\sum_{\rm  even}\Big(E_{\bar\varepsilon},E_{-\bar\varepsilon} (~,~) \Big)_{\al} = - 
\sum_{\rm  odd}\Big(~,~(E_{\bar\varepsilon},E_{-\bar\varepsilon} ) \Big)_{\al+\frac12}\ ,\\
\sum_{\rm  odd}\Big(E_{\bar\varepsilon},E_{-\bar\varepsilon} (~,~) \Big)_{\al} = +
\sum_{\rm  even}\Big(~,~(E_{\bar\varepsilon},E_{-\bar\varepsilon} ) \Big)_{\al+\frac12}\ .
\end{align}
Similar relations hold for $\varepsilon$.

\item 
Identities without sums:

\bea
\label{SU(4)id}
\frac{\p}{\p\eta^{[2-2\al]}}\frac{\p}{\p\zeta^{[2+2\al]}}\left[
\left(\eta^m\frac{\p}{\p\eta^n}+\zeta^m\frac{\p}{\p\zeta^n}\right)
-\qter\da^m{}_n\left(\eta^k\frac{\p}{\p\eta^k}+\zeta^k\frac{\p}{\p\zeta^k}\right)\right]=0
\eea
and
\bea
\label{inoutID}
&&
\frac{d^{[4]}}{2\p^{+2}}
\left[ \frac{\p}{\p\eta^{[2-2\al]}}\frac{\p}{\p\zeta^{[2+2\al]}}\cK_\al^{a\,(0,\eta,\zeta)} \right]^\ast \nn\\
&=& \frac{\p}{\p\eta^{[2+2\al]}}\frac{\p}{\p\zeta^{[2-2\al]}}
\frac{(-1)^{2\al}}{\p^{+2\al}}\left(\p^{+2\al},\frac{1}{\p^{+(-4\al)}}\Big(\p^{+(-2\al)},\p^{+(-2\al)}\Big)\right)
\cK_\al^{a\,(0,\eta,\zeta)},
\eea
which are valid for each $\al \in \{-1,-1/2,0,+1/2,+1\}$ after setting $\eta=\zeta=0$.

\item The Recursion Relation \\

\begin{align}
\d^{+}\Big(\frac{1}{\d^{+}},\frac{1}{\d^{+2}}(\d^{+},\d^{+})\Big)_{\al} =
\Big(~,~(~,~)\Big)_{\al+1} \ ,
\end{align}
is derived using,
\bea
\label{TbarID}
\frac{\p}{\p\eta^{[2-2\al]}}\frac{\p}{\p\zeta^{[2+2\al]}}\om^{m n}\frac{\p}{\p\eta^m}\frac{\p}{\p\eta^n}
&=& \frac{\p}{\p\eta^{[4-2\al]}}\frac{\p}{\p\zeta^{[+2\al]}}\om^{m n}\frac{\p}{\p\zeta^m}\frac{\p}{\p\zeta^n}
\eea
and
\bea
\label{Tid}
\frac{\p}{\p\eta^{[2-2\al]}}\frac{\p}{\p\zeta^{[2+2\al]}}\ombar_{m n}\eta^m\eta^n
&=& -2\ombar_{[2]}\frac{\p}{\p\eta^{[-2\al]}}\frac{\p}{\p\zeta^{[2+2\al]}}
\quad\text{(or $=0$)}
 \nn\\
\frac{\p}{\p\eta^{[2-2\al]}}\frac{\p}{\p\zeta^{[2+2\al]}}\ombar_{m n}\zeta^m\zeta^n
&=& -2\ombar_{[2]}\frac{\p}{\p\eta^{[2-2\al]}}\frac{\p}{\p\zeta^{[+2\al]}},
\quad\text{(or $=0$)}
\eea
where we defined ($m=0,2$)
\bea
A_{[m]}B_{[n]}C_{[4-m-n]} \equiv \frac{1}{m!n!(4-m-n)!}\eps^{i_1\dots i_4}
A_{i_1\dots i_m}B_{i_{m+1}\dots i_{m+n}}C_{i_{m+n+1}\dots i_4}.
\eea
The identity (\ref{Tid}) is valid only after setting $\eta=\zeta=0$, and its right hand side vanishes whenever the power of the $\eta$- or $\zeta$-derivative there comes out to be negative.

Using (\ref{TbarID}) together with shifting $\al\rightarrow\al+1$ to bring the sums to common limits, we find
\bea
\label{Trec1}
\esum\left(\frac{\p}{\p\eta^{m n}}\mc{S}+\frac{\p}{\p\zeta^{m n}}\mc{T} \right)
\cK_\al^{a\,(\eps,\eta,\zeta)}
=\sum_{\rm{even}}^{\al\neq +1}\frac{\p}{\p\zeta^{m n}}\Big(
-\mc{S}\cK_{\al+1}^{a\,(\eps,\eta,\zeta)}+T\cK_\al^{a\,(\eps,\eta,\zeta)} \Big).
\eea
In a similar way, (\ref{Tid}) implies
\bea
\label{Trec2}
&& \ombar_{m n}\esum\Big(\eta^m\eta^n\mc{S}^{-1}+\zeta^m\zeta^n\mc{T}^{-1}\Big)\cK_\al^{a\,(\eps,\eta,\zeta)}, \nn\\
&=& -2\ombar_{[2]}\sum_{\al = -1,0}(-1)^\al\frac{\p}{\p\eta^{[-2\al]}}\frac{\p}{\p\zeta^{[2+2\al]}}
\Big(\mc{S}^{-1}\cK_\al^{a\,(\eps,\eta,\zeta)}-\mc{T}^{-1}\cK_{\al+1}^{a\,(\eps,\eta,\zeta)}\Big).
\eea
It is then obvious that the vanishing of (\ref{Trec1}) and (\ref{Trec2}) requires the recursion relation (\ref{rec1}). The proof in the odd case is similar.

\item {Inside-Out-Constraint}\\

Finally, the identity (\ref{inoutID}) implies that, in the even case,

\bea
\frac{d^{[4]}}{2\p^{+2}}\left(\esum\cK_\al^{a\,(\eps,\eta,\zeta)}\right)^\ast
=\esum \frac{1}{\p^{+(-2\al)}}\left(\p^{+(-2\al)},\frac{1}{\p^{+4\al}}\Big(\p^{+2\al},\p^{+2\al}\Big)\right)
\cK_{-\al}^{a\,(\epsbar,\eta,\zeta)}.
\eea
The inside-out constraint requires 

\bea
\frac{d^{[4]}}{2\p^{+2}}\left(\esum\cK_\al^{a\,(\eps,\eta,\zeta)}\right)^\ast
=\esum\cK_\al^{a\,(\epsbar,\eta,\zeta)},
\eea
which demands the following relations between the exponents
\bea
&&
A_\al=A_{-\al}-2\al, \quad
B_\al=B_{-\al}-2\al, \quad
M_\al=M_{-\al}+4\al ,\nn\\
&&
C_\al=C_{-\al}+2\al, \quad
D_\al=D_{-\al}+2\al.
\eea
These relations, in turn,  follow from the recursion relation (\ref{rec2}). For example,
\bea
A_{\al+1}=A_\al-1 \quad\Rightarrow\quad A_{\al+k}=A_\al-k \quad\Rightarrow\quad A_{-\al}=A_\al-(-2\al).
\eea
Therefore, the recursion relation (\ref{rec2}) implies that the inside-out constraint is satisfied. The same is true in the odd case.

\end{itemize}


\setcounter{equation}{0}
\section{Calculating $\dab_{\mc{P}^{-}}^{(1)}\phi^a$ and $\dab_{\mc{J}^{-}}^{(1)}\phi^a$ }\label{App:P-J-}

The first commutator in (\ref{QQisP1}) involves, in the odd case,
\bea
\dab_{\ep\Qbar}^{free}\phi^a=\frac{1}{\sqrt2}\ep^m\qbar_m\frac{\p}{\p^{+}}\phi^a, \quad
\dab_{\epbar\cQ}^{int}\phi^a=i\epbar_n\mc{S}^{-1}\osum\eta^n\cK_\al^{a\,(0,\eta,\zeta)}.
\eea
Applying the master formula (\ref{master}) with $\mc{O}=\qbar_m\p/\p^{+}$, we note that all the commutators vanish, whereas for the triplets the formula (\ref{D1D2}) can be applied. This gives~\footnote{
Setting $r=0$ after the differentiation is kept implicit.
}
\bea
[\dab_{\ep\Qbar}^{free},\dab_{\epbar\cQ}^{int}]\phi^a=\frac{i}{\sqrt2}\ep^m\epbar_n\mc{S}^{-1}\osum
\eta^n\left(\frac{\p}{\p\eta^m}\frac{\p}{\p r}\mc{S} K_\al^{a\,[r,1]}
+\frac{\p}{\p\zeta^m}\frac{\p}{\p r}\mc{T} K_\al^{a\,[1,r]}\right),
\eea
where (\ref{qisd}) has also been used. The second commutator in (\ref{QQisP1}) involves
\bea
\dab_{\epbar\cQ}^{free}\phi^a=i\epbar_n\ta^n\p\phi^a, \quad
\dab_{\ep\Qbar}^{int}\phi^a=\frac{1}{\sqrt2}\ep^m\osum\frac{\p}{\p\eta^m}\cK_\al^{a\,(0,\eta,\zeta)}.
\eea
With $\mc{O}=\ta^n\p$, both triplets in the master formula (\ref{master}) vanish. Using
\bea
[E_\eta,\ta^n\p]=\eta^n\wh\p E_\eta
\eea
and the identity (\ref{doublet}), we find
\bea
[\dab_{\epbar\cQ}^{free},\dab_{\ep\Qbar}^{int}]\phi^a=\frac{i}{\sqrt2}\epbar_n\ep^m\osum
\frac{\p}{\p\eta^m}\frac{\p}{\p r}\Big(\eta^n K_\al^{a\,[r,1]}
+\zeta^n K_\al^{a\,[1,r]}\Big).
\eea
Therefore,
\bea
[\dab_{\ep\Qbar}^{free},\dab_{\epbar\cQ}^{int}]\phi^a-
[\dab_{\epbar\cQ}^{free},\dab_{\ep\Qbar}^{int}]\phi^a
=\frac{i}{\sqrt2}\ep^m\epbar_n\frac{\p}{\p r}\osum\left(
\eta^n\frac{\p}{\p\eta^m}+\frac{\p}{\p\eta^m}\eta^n\right) K_\al^{a\,[r,1]} \nn\\
+\frac{i}{\sqrt2}\ep^m\epbar_n\frac{\p}{\p r}\osum\left(
\eta^n\frac{\p}{\p\zeta^m} K_{\al+1}^{a\,[1,r]}
-\zeta^n\frac{\p}{\p\eta^m} K_\al^{a\,[1,r]}\right).
\eea
Using the identity (B.19), this becomes
\bea
[\dab_{\ep\Qbar}^{free},\dab_{\epbar\cQ}^{int}]\phi^a-
[\dab_{\epbar\cQ}^{free},\dab_{\ep\Qbar}^{int}]\phi^a
=\frac{i}{\sqrt2}\ep^m\epbar_m\frac{\p}{\p r}\left(
\osum K_\al^{a\,[r,1]}
+\esum K_{\al+\half}^{a\,[1,r]}\right)
\eea
so that the $O(f)$ part of the Hamiltonian shift in the odd case is 
\bea
\label{P-odd2}
\dab_{\mc{P}^{-}}^{(1)\,\rm odd}\phi^a=-\frac{i}{2}\frac{\p}{\p r}\left(
\osum K_\al^{a\,[r,1]}+\esum K_{\al+\half}^{a\,[1,r]}\right).
\eea
which reproduces (\ref{P-odd}). In the even case, because of (B.18), the corresponding expression has the relative \emph{minus} sign, which explains the rule (\ref{oddtoeven}).

The $O(f)$ part of the Lorentz boost follows from commuting with the kinematical special conformal transformation $K$,
\bea
[\da_K,\dab_{\mc{P}^{-}}^{(1)}]\phi^a=2i\dab_{\mc{J}^{-}}^{(1)}\phi^a,
\eea
with $\da_K\phi^a=2i x\mc{A}\phi^a$. Defining $\da_{x\mc{A}}\phi^a=x\mc{A}\phi^a$, we have
\bea
\dab_{\mc{J}^{-}}^{(1)\,\rm odd}\phi^a=-\frac{i}{2}\frac{\p}{\p r}\left(
\osum\Big[\da_{x\mc{A}},K_\al^{a\,[r,1]}\Big]
+\esum\Big[\da_{x\mc{A}},K_{\al+\half}^{a\,[1,r]}\Big]\right).
\eea
In calculating the commutator part of the master formula (\ref{master}), we use
\bea
[E_r E_\eta\p^{+k},x\mc{A}]=\left(x\left[\half\eta\frac{\p}{\p\eta}-k\right]
+r x^{-}+r\left[\half\left(\eta\frac{\p}{\p\eta}-\mc{N}\right)+\frac{3}{2}-k\right]\frac{1}{\p^{+}}\right)
E_r E_\eta \p^{+k}. \quad
\eea
The $x^{-}$-dependent contributions cancel because $E_r$ comes with $E_{-r}$. For the $x$-dependent contributions, we note that moving $x$ to the left involves
\bea
[E_r,x]=r\frac{1}{\p^{+}}E_r.
\eea
As $x\mc{A}$ is a derivative operator \emph{plus} $x/2$, the contribution from the triplets in (\ref{master}) is
\bea
\Big(\Big\{x\mc{A}\Big\}_{12}+\Big\{x\mc{A}\Big\}_{34}\Big) K_\al^{a\,[r,1]}
&=& \left(x-\frac{r}{2}\wh{\mc{U}}_2\right) K_\al^{a\,[r,1]},
\nn\\
\Big(\Big\{x\mc{A}\Big\}_{12}+\Big\{x\mc{A}\Big\}_{34}\Big) K_\al^{a\,[1,r]}
&=& x K_\al^{a\,[1,r]}.
\eea
Alltogether, we find
\bea
\Big[\da_{x\mc{A}},K_\al^{a\,[r,1]}\Big]
&=& \Big\{-x+\frac{r}{2}(\mc{O}_\eta\wh{\mc{U}}_1-\mc{O}_\eta\wh{\mc{U}}_2)
+r\left(\frac{3}{2}-B_\al\right)\wh{\mc{U}}_1, \nn\\
&&\hspace{153pt}
-r\left(M_\al-C_\al-D_\al+3+\al\right)\wh{\mc{U}}_2\Big\} K_\al^{a\,[r,1]}
\nn\\[3pt]
\Big[\da_{x\mc{A}},K_\al^{a\,[1,r]}\Big],
&=& \Big\{ -x+\frac{r}{2}(\mc{O}_\zeta\wh{\mc{U}}_3-\mc{O}_\zeta\wh{\mc{U}}_4)
+r\left(\frac{3}{2}-C_\al\right)\wh{\mc{U}}_3-r\left(\frac{3}{2}-D_\al\right)\wh{\mc{U}}_4\Big\}
K_\al^{a\,[1,r]}, \nn\\
\eea
where we defined
\bea
\label{Oeta}
\mc{O}_\eta \equiv \eta\frac{\p}{\p\eta}-\mc{N}, \quad
\mc{O}_\zeta \equiv \zeta\frac{\p}{\p\zeta}-\mc{N}
\eea
and used that, when $\eta$- and $\zeta$-derivatives act on the \emph{whole} $K_\al^a\equiv\cK_\al^{a\,(0,\eta,\zeta)}$, we have
\bea
\eta\frac{\p}{\p\eta}=2-2\al, \quad
\zeta\frac{\p}{\p\zeta}=2+2\al
\eea
according to the definition of the sums in (\ref{esum}) and (\ref{osum}). It then follows that~\footnote{
Setting $u=0$ after the differentiation is kept implicit.  }
\bea
\label{J-odd}
\dab_{\mc{J}^{-}}^{(1)\,\rm odd}\phi^a &=& -x\dab_{\mc{P}^{-}}^{(1)\,\rm odd}\phi^a
-\frac{i}{2}\Bigg\{\half\frac{\p}{\p u}\left(
\osum K_\al^{a\,\{u,1\}}+\esum K_{\al+\half}^{a\,\{1,u\}}\right) \nn\\[5pt]
&&
-\osum\left[ \left(B_\al-\frac{3}{2}\right)\wh{\mc{U}}_1+(M_\al-C_\al-D_\al+3+\al)\wh{\mc{U}}_2 \right] 
K_\al^a \nn\\
&&
-\esum\left[ \left(C_{\al+\half}-\frac{3}{2}\right)\wh{\mc{U}}_3
-\left(D_{\al+\half}-\frac{3}{2}\right)\wh{\mc{U}}_4 \right]
K_{\al+\half}^a, \Bigg\}
\eea
where we defined
\bea
\label{ueta}
K_\al^{a\,\{u,1\}} \equiv (E_{u,\eta}\mc{U}_1)(E_{-u,\eta}\mc{U}_2) K_\al^a, \quad
K_\al^{a\,\{1,u\}} \equiv (E_{u,\zeta}\mc{U}_3)(E_{-u,\zeta}\mc{U}_4) K_\al^a
\eea
with
\bea
\label{Eueta}
E_{u,\eta} \equiv e^{u\wh{\mc{O}}_\eta}, \quad
E_{u,\zeta} \equiv e^{u\wh{\mc{O}}_\zeta}.
\eea
The result in the even case is obtained by the substitution (\ref{oddtoeven}).

\setcounter{equation}{0}
\section{Calculating $[\dab_{\mc{P}^{-}},\dab_{\mc{J}^{-}}]\phi^a$ }\label{App:[P-,J-]}

The commutator of the Hamiltonian shift $\dab_{\mc{P}^{-}}\phi^a$ with the Lorentz boost $\dab_{\mc{J}^{-}}\phi^a$ is
\bea
\label{PJcomm}
[\dab_{\mc{P}^{-}},\dab_{\mc{J}^{-}}]\phi^a=
[\dab_{\mc{P}^{-}}^{free},\dab_{\mc{J}^{-}}^{(1)}]\phi^a+[\dab_{\mc{P}^{-}}^{(1)},\dab_{\mc{J}^{-}}^{free}]\phi^a
+O(f^2),
\eea
where
\bea
\dab_{\mc{P}^{-}}^{free}\phi^a=-\frac{i}{2}\frac{\p^2}{\p^{+}}\phi^a, \quad
\dab_{\mc{J}^{-}}^{free}\phi^a=-i\frac{\p}{\p^{+}}\mc{A}\phi^a
\eea
and $\dab_{\mc{P}^{-}}^{(1)}\phi^a$ with $\dab_{\mc{J}^{-}}^{(1)}\phi^a$, in the odd case, are given in (\ref{P-odd2}) and (\ref{J-odd}). 

In $[\dab_{\mc{P}^{-}}^{free},\dab_{\mc{J}^{-}}^{(1)}]\phi^a$, only the the first term in (\ref{J-odd}), with explicit $x$, contibutes to the commutator part of the master formula (\ref{master}). For the triplets in (\ref{master}), we can use (\ref{D1D2}) which gives
\bea
[\da_{\mc{P}^{-}}^{free},K_\al^a]=-\frac{i}{2}\mc{S}
\frac{\p^2}{\p r\p r^\prime}\left(
K_\al^{a\,[r+r^\prime,1]}
+K_{\al+1}^{a\,[1,r+r^\prime]} \right),
\eea
where also the recursion relation (\ref{rec1}) has been used. It then immediately follows that
\bea
\label{PonJ}
[\dab_{\mc{P}^{-}}^{free},\dab_{\mc{J}^{-}}^{(1)\,\rm odd}]\phi^a 
&=& -x[\dab_{\mc{P}^{-}}^{free},\dab_{\mc{P}^{-}}^{(1)\,\rm odd}]\phi^a
-i\frac{\p}{\p^{+}}\dab_{\mc{P}^{-}}^{(1)\,\rm odd}\phi^a 
-\frac{1}{4}\mc{S}\frac{\p^2}{\p r\p r^\prime}
\Bigg\{ \nn\\
&&\hspace{-80pt}
-\osum\left[ \left(B_\al-\frac{3}{2}\right)\wh{\mc{U}}_1+(M_\al-C_\al-D_\al+3+\al)\wh{\mc{U}}_2 \right] 
\left( K_\al^{a\,[r+r^\prime,1]}+K_{\al+1}^{a\,[1,r+r^\prime]} \right) \nn\\
&&\hspace{-80pt}
-\esum\left[ \left(C_{\al+\half}-\frac{3}{2}\right)\wh{\mc{U}}_3
-\left(D_{\al+\half}-\frac{3}{2}\right)\wh{\mc{U}}_4 \right]
\left( K_{\al+\half}^{a\,[r+r^\prime,1]}
+K_{\al+\frac{3}{2}}^{a\,[1,r+r^\prime]} \right) \nn\\[5pt]
&&\hspace{-125pt}
+\half\frac{\p}{\p u}\Bigg[
\osum\Big(K_\al^{a\,[r+r^\prime,1]\{u,1\}}+K_{\al+1}^{a\,[1,r+r^\prime]\{u,1\}}\Big) 
+\esum\Big(K_{\al+\half}^{a\,[r+r^\prime,1]\{1,u\}}+K_{\al+\frac{3}{2}}^{a\,[1,r+r^\prime]\{1,u\}}\Big)
\Bigg]
\Bigg\}.
\eea

The commutator $[\dab_{\mc{J}^{-}}^{free},\dab_{\mc{P}^{-}}^{(1)}]\phi^a$ requires longer analysis. First, we apply the master formula (\ref{master}) noting that
\bea
\label{comJP}
[E_r E_\eta\p^{+k},\frac{\p}{\p^{+}}\mc{A}]=\frac{\p}{\p^{+}}\left(
\half r\frac{\p}{\p r}+\half\eta\frac{\p}{\p\eta}-k\right)E_r E_\eta\p^{+k}.
\eea
For the triplets in (\ref{master}), we have
\bea
\label{tripJP}
\Big\{\frac{\p}{\p^{+}}\mc{A}\Big\}
=\Big\{(\mc{A}+\half)\frac{\p}{\p^{+}}\Big\}
=-\half\Big\{x\frac{\p^2}{\p^{+}}\Big\}-\half\Big\{\mc{N}\wh\p\Big\}+\Big\{\wh\p\Big\},
\eea
where the $x^{-}$-dependent part dropped out. For the $x$-dependent part, we observe that
\bea
\left( \Big\{x\frac{\p^2}{\p^{+}}\Big\}_{12}+\Big\{x\frac{\p^2}{\p^{+}}\Big\}_{34} \right)K_\al^{a\,[r,1]}
&=& 2i x\left[\dab_{\mc{P}^{-}}^{free},K_\al^{a\,[r,1]}\right]
-r\wh{\mc{U}}_2 \Big\{\frac{\p^2}{\p^{+}}\Big\}_{34} K_\al^{a\,[r,1]} \nn\\
\left( \Big\{x\frac{\p^2}{\p^{+}}\Big\}_{12}+\Big\{x\frac{\p^2}{\p^{+}}\Big\}_{34} \right)K_\al^{a\,[1,r]}
&=& 2i x\left[\dab_{\mc{P}^{-}}^{free},K_\al^{a\,[1,r]}\right],
\eea
where, using (\ref{D1D2}) and the recursion relation (\ref{rec1}), we also have
\bea
\Big\{\frac{\p^2}{\p^{+}}\Big\}_{34} K_\al^{a\,[r,1]}
=\frac{\p}{\p r^\prime}\frac{\p}{\p r^{\prime\prime}}\mc{S}
K_{\al+1}^{a\,[r,r^\prime+r^{\prime\prime}]}.
\eea
For the $\eta\frac{\p}{\p\eta}$ and $\zeta\frac{\p}{\p\zeta}$ parts arising from (\ref{comJP}), we use
\bea
\eta\frac{\p}{\p\eta}\wh\p\mc{U}_1+\eta\frac{\p}{\p\eta}\wh\p\mc{U}_2
&=& \Big\{\eta\frac{\p}{\p\eta}\wh\p\Big\}_{12}+(2-2\al)\wh\p, \nn\\
\zeta\frac{\p}{\p\zeta}\wh\p\mc{U}_3+\zeta\frac{\p}{\p\zeta}\wh\p\mc{U}_4
&=& \Big\{\zeta\frac{\p}{\p\zeta}\wh\p\Big\}_{34}+(2+2\al)\wh\p\mc{U}_2.
\eea
These triplets then combine with the $\mc{N}$-dependent triplets in (\ref{tripJP}), so that the combinations (\ref{Oeta}) form, and we can use
\bea
\Big\{\mc{O}_\eta\wh\p\Big\}_{12} K_\al^a
=\mc{S}\frac{\p}{\p u}\frac{\p}{\p r^\prime} K_\al^{a\,[r^\prime,1]\{u,1\}}, \quad
\Big\{\mc{O}_\zeta\wh\p\Big\}_{34} K_\al^a
=\mc{S}\frac{\p}{\p u}\frac{\p}{\p r^\prime} K_{\al+1}^{a\,[1,r^\prime]\{1,u\}}.
\eea
Finally, using the following ``recombination'' identity,
\bea
&&\Big( \la_1\wh\p\mc{U}_1+\la_2\wh\p\mc{U}_2+\la_3\wh\p\mc{U}_3+\la_4\wh\p\mc{U}_4 \Big)
K_\al^a \;=\; (\la_1+\la_2+\la_3+\la_4)\wh\p K_\al^a \nn\\
&+&\mc{S}\frac{\p}{\p r^\prime}\left\{
\Big[ \la_1\wh{\mc{U}}_1-(\la_2+\la_3+\la_4)\wh{\mc{U}}_2 \Big] K_\al^{a\,[r^\prime,1]}
+\Big(\la_3\wh{\mc{U}}_3-\la_4\wh{\mc{U}}_4\Big) K_{\al+1}^{a\,[1,r^\prime]} \right\}
\eea 
we obtain
\bea
\label{JonP}
[\dab_{\mc{J}^{-}}^{free},\dab_{\mc{P}^{-}}^{(1)\,\rm odd}]\phi^a
&=& -x[\dab_{\mc{P}^{-}}^{free},\dab_{\mc{P}^{-}}^{(1)\,\rm odd}]\phi^a
-i\frac{\p}{\p^{+}}\dab_{\mc{P}^{-}}^{(1)\,\rm odd}\phi^a
-\frac{1}{4}\mc{S}\frac{\p^2}{\p r\p r^\prime}\Bigg\{ 
\osum\wh{\mc{U}}_2 K_{\al+1}^{a\,[1,r+r^\prime]} \nn\\
&&\hspace{-40pt}
+\osum\Big[\Big(3-2B_\al\Big)\wh{\mc{U}}_1-\Big(2(M_\al-C_\al-D_\al)+7+2\al\Big)\wh{\mc{U}}_2\Big]
K_\al^{a\,[r+r^\prime,1]} \nn\\
&&\hspace{-40pt}
+\osum\Big[\Big(2-2C_\al\Big)\wh{\mc{U}}_3-\Big(2-2D_\al\Big)\wh{\mc{U}}_4\Big]
K_{\al+1}^{a\,[r,r^\prime]} \nn\\
&&\hspace{-70pt}
+\esum\Big[\Big(2-2B_{\al+\half}\Big)\wh{\mc{U}}_1
-\Big(2(M_{\al+\half}-C_{\al+\half}-D_{\al+\half})+8+2\al\Big)\wh{\mc{U}}_2\Big]
K_{\al+\half}^{a\,[r,r^\prime]} \nn\\
&&\hspace{-70pt}
+\esum\Big[\Big(3-2C_{\al+\half}\Big)\wh{\mc{U}}_3-\Big(3-2D_{\al+\half}\Big)\wh{\mc{U}}_4\Big]
K_{\al+\frac{3}{2}}^{a\,[1,r+r^\prime]} \nn\\
&&\hspace{-100pt}
+\frac{\p}{\p u}\left[
\osum\Big(
K_\al^{a\,[r+r^\prime,1]\{u,1\}}
+K_{\al+1}^{a\,[r,r^\prime]\{1,u\}}\Big)
+\esum\Big(
K_{\al+\half}^{a\,[r,r^\prime]\{u,1\}}
+K_{\al+\frac{3}{2}}^{a\,[1,r+r^\prime]\{1,u\}}\Big)\right] \Bigg\}.
\eea

The $O(f)$ part of (\ref{PJcomm}) is the difference of (\ref{PonJ}) and (\ref{JonP}). We see that the first two terms on the right hand side of (\ref{JonP}) cancel the corresponding terms in (\ref{PonJ}).
Using the following identities,
\bea
\frac{\p}{\p u}\Big(\esum K_\al^{a\,\{u,1\}}+\osum K_{\al+\half}^{a\,\{1,u\}}\Big)
&=& 4\esum\wh{\mc{U}}_2 K_\al^a
+\esum(2-2\al)(\wh{\mc{U}}_1-\wh{\mc{U}}_2) K_\al^a \nn\\
&&\hspace{60pt}
+\osum(2+2\al)(\wh{\mc{U}}_3-\wh{\mc{U}}_4) K_{\al+\half}^a
\nn\\
\frac{\p}{\p u}\Big(\osum K_\al^{a\,\{u,1\}}-\esum K_{\al+\half}^{a\,\{1,u\}}\Big)
&=& 4\osum\wh{\mc{U}}_2\cK_\al^a
+\osum(2-2\al)(\wh{\mc{U}}_1-\wh{\mc{U}}_2) K_\al^a \nn\\
&&\hspace{60pt}
-\esum(2+2\al)(\wh{\mc{U}}_3-\wh{\mc{U}}_4) K_{\al+\half}^a
\eea
we find that the contribution to the difference of (\ref{PonJ}) and (\ref{JonP}) from the terms with $\frac{\p}{\p u}$ is
\bea
-\frac{1}{4}\mc{S}\frac{\p^2}{\p r\p r^\prime}\Bigg\{
-\osum\Big[(1-\al)\wh{\mc{U}}_1+(1+\al)\wh{\mc{U}}_2\Big] K_\al^{a\,[r+r^\prime,1]}
+\esum(1+\al)\Big(\wh{\mc{U}}_3-\wh{\mc{U}}_4\Big) K_{\al+\half}^{a\,[r+r^\prime,1]} \nn\\
+\osum\Big[(1-\al)\wh{\mc{U}}_1+(1+\al)\wh{\mc{U}}_2\Big] K_{\al+1}^{a\,[1,r+r^\prime]}
-\esum(1+\al)\Big(\wh{\mc{U}}_3-\wh{\mc{U}}_4\Big) K_{\al+\frac{3}{2}}^{a\,[1,r+r^\prime]} \nn\\
-2\esum\Big[(1-\al)\wh{\mc{U}}_1+(1+\al)\wh{\mc{U}}_2\Big] K_{\al+\half}^{a\,[r,r^\prime]}
-2\osum(1+\al)\Big(\wh{\mc{U}}_3-\wh{\mc{U}}_4\Big) K_{\al+1}^{a\,[r,r^\prime]}\Bigg\}.
\hspace{20pt}
\eea
The complete expression for the difference of (\ref{PonJ}) and (\ref{JonP}) then easily follows, and we find that the exponents $B_\al$, $M_\al$, $C_\al$, $D_\al$ appear in combinations with explicit $\al$'s that are \emph{$\al$-independent} thanks to the recursion relations (\ref{rec2}). The final answer for the $O(f)$ part of the commutator (\ref{PJcomm}) is given in equations (\ref{PJodd}) through (\ref{O12}).


\section{Odd Ansatz: the BLG Solution}\label{App:odd}
\setcounter{equation}{0}  
As we mentioned in the text, the commutator $[\, \deltab_{\m P^{-}}\,,\, \deltab_{\m J^{-}}\,]\,\varphi^{a}$  contains two powers of the transverse derivative and four powers of $\bar d$. A necessary condition for its vanishing is that the terms for which $\partial^2\bar d^{[4]}$ act on the same superfield vanish by themselves. In this appendix, we single those terms out for the odd Ansatz. 

There are two terms where  $\partial^2\bar d^{[4]}$ acts on the ``first" superfield, 
\begin{align}
(C_{-\frac12}-\frac32)\frac{1}{\d^{+}} \Big( \frac{\d^{2}\bar d^{[4]}}{\d^{+5}}, \d^{+}(\frac{1}{\d^{+}},~) \Big) 
- (D_{-\frac12}-\frac32)\frac{1}{\d^{+}} \Big( \frac{\d^{2}\bar d^{[4]}}{\d^{+5}}, \d^{+}(~,\frac{1}{\d^{+}}) \Big)\  ,\label{firstE} 
\end{align} 
which must be cancelled by terms where $\partial^2\bar d^{[4]}$ acts on the ``second" and ``third" superfields. If we require that the terms  with the most inverse powers of delplus (most singular) vanish by themselves, we arrive at  

\begin{align}\label{MMM-BECM}
M_{-\frac12} \,=\, 2(C_{-\frac12}\,+\,D_{-\frac12})\,-\,6\,\le\, -\,2\ ,\quad  B_{-\frac12}\,=\,C_{-\frac12}\,-\,M_{-\frac12}\ .
\end{align}
These terms reduce to (dropping the subscript ${-\frac12}$) and we will write $\bar d^{[4]}$ as $\bar d^4$ in the remaining calculations in this and the remaining appendices to get the expressions more transparent.

\begin{align}\label{remain}
&2(B-3) \bigg[ -2\Big(\frac{1}{\d^{+2}},\frac{1}{\d^{+2}}(~,\frac{\d^{2}\dbar^{4}}{\d^{+2}})\Big)+\d^{+}\Big(\frac{1}{\d^{+2}},\frac{1}{\d^{+2}}(~,\frac{\d^{2}\dbar^{4}}{\d^{+3}})\Big)+\frac{1}{\d^{+}}\Big(\frac{1}{\d^{+2}},\frac{1}{\d^{+2}}(~,\frac{\d^{2}\dbar^{4}}{\d^{+}})\Big)\bigg]\cr
&-(C+D-3)\bigg[-\d^{+2}\Big(\frac{1}{\d^{+2}},\frac{1}{\d^{+2}}(~,\frac{\d^{2}\dbar^{4}}{\d^{+4}})\Big) +3 \d^{+}\Big(\frac{1}{\d^{+2}},\frac{1}{\d^{+2}}(~,\frac{\d^{2}\dbar^{4}}{\d^{+3}})\Big)\cr
&\qquad\qquad-3\Big(\frac{1}{\d^{+2}},\frac{1}{\d^{+2}}(~,\frac{\d^{2}\dbar^{4}}{\d^{+2}})\Big)+\frac{1}{\d^{+}}\Big(\frac{1}{\d^{+2}},\frac{1}{\d^{+2}}(~,\frac{\d^{2}\dbar^{4}}{\d^{+}})\Big)\bigg]\cr
&+(C-\frac32)\bigg[-\d^{+2}\Big(\frac{1}{\d^{+2}},\frac{1}{\d^{+}}(\frac{1}{\d^{+}},\frac{\d^{2}\dbar^{4}}{\d^{+4}})\Big)+3\d^{+}\Big(\frac{1}{\d^{+2}},\frac{1}{\d^{+}}(\frac{1}{\d^{+}},\frac{\d^{2}\dbar^{4}}{\d^{+3}})\Big)\cr
&\qquad\qquad-3\Big(\frac{1}{\d^{+2}},\frac{1}{\d^{+}}(\frac{1}{\d^{+}},\frac{\d^{2}\dbar^{4}}{\d^{+2}})\Big)+\frac{1}{\d^{+}}\Big(\frac{1}{\d^{+2}},\frac{1}{\d^{+}}(\frac{1}{\d^{+}},\frac{\d^{2}\dbar^{4}}{\d^{+}})\Big)\cr
&
+2\d^{+}\Big(\frac{1}{\d^{+}},\frac{1}{\d^{+2}}(\frac{1}{\d^{+}},\frac{\d^{2}\dbar^{4}}{\d^{+3}})\Big)
-4\Big(\frac{1}{\d^{+}},\frac{1}{\d^{+2}}(\frac{1}{\d^{+}},\frac{\d^{2}\dbar^{4}}{\d^{+2}})\Big)
+2\frac{1}{\d^{+}}\Big(\frac{1}{\d^{+}},\frac{1}{\d^{+2}}(\frac{1}{\d^{+}},\frac{\d^{2}\dbar^{4}}{\d^{+}})\Big)\bigg]\ ,
\end{align}
where we used

\begin{align}
&(C-3/2)(\frac{1}{\partial^+}\,,\,\frac{\partial^2\bar d^4}{\partial^{+a}})-(D-3/2)(~\,,\,\frac{\partial^2\bar d^4}{\partial^{+(a+1)}}\,)~=~\nonumber\\
&(C-3/2)\partial^+(\frac{1}{\partial^+}\,,\,\frac{\partial^2\bar d^4}{\partial^{+(a+1)}})-(C+D-3)(~\,,\,\frac{\partial^2\bar d^4}{\partial^{+(a+1)}}\,)\ .
\end{align}
One then sees that many terms can vanish due to antisymmetry. For example, the first three lines in \eqref{remain} vanish as long as $f^{a}{}_{bcd}=-f^{a}{}_{cbd}$ and  $M=-2$.

In order to investigate the possible cases for $M\leq -2$, we set $M=-2-m$ ($m\ge 0$) and  $B=C-M=C+2-m$. 
When $m=0$, \eqref{remain} is rewritten as

\begin{align}
&-(C-\frac{3}{2})\frac{f^{a}{}_{bcd}}{\partial^{+(A+1)}}\,\cdot  \nn\\
&\bigg[ \,\partial^{+3}\,(\partial^{+C}\varphi^b\,\partial^{+(C-1)}\varphi^c\,\partial^2\bar d^4\partial^{+(D-3)}\varphi^d\,)
-3\,\partial^{+2}\,(\partial^{+C}\varphi^b\,\partial^{+(C-1)}\varphi^c\,\partial^2\bar d^4\partial^{+(D-2)}\varphi^d\,) \nn \\
&
+3\,\partial^{+}\,(\partial^{+C}\varphi^b\,\partial^{+(C-1)}\varphi^c\,\partial^2\bar d^4\partial^{+(D-1)}\varphi^d\,)
-(\partial^{+C}\varphi^b\,\partial^{+(C-1)}\varphi^c\,\partial^2\bar d^4\partial^{+D}\varphi^d\,) \nn\\
&
-2\,\partial^{+2}\,(\partial^{+(C+1)}\varphi^b\,\partial^{+(C-1)}\varphi^c\,\partial^2\bar d^4\partial^{+(D-3)}\varphi^d\,)
+4\,\partial^{+}\,(\partial^{+(C+1)}\varphi^b\,\partial^{+(C-1)}\varphi^c\,\partial^2\bar d^4\partial^{+(D-2)}\varphi^d\,)\nn\\
&
-2\,(\partial^{+(C+1)}\varphi^b\,\partial^{+(C-1)}\varphi^c\,\partial^2\bar d^4\partial^{+(D-1)}\varphi^d\,)\bigg]\ ,
\end{align}
which can be reorganized along terms of the form $\partial^2\bar d^4\partial^{+(D-n)}\varphi^d$ ($n=1,2,3$), yielding that all cancel, except for

\begin{align}
\label{remaining}
-\,(C-\frac32) f^{a}{}_{bcd}\Big[(\partial^2\bar d^4\partial^{+(D-3)}\varphi^d)\,\partial^{+2}(\partial^{+(C+1)}\varphi^b\,\partial^{+(C-1)}\varphi^c)\Big] \ .
\end{align}
We then compare this to \eqref{firstE} when $B=C+2$, which is 

\begin{align}
+(C-\frac32) f^{a}{}_{bcd}\Big[(\partial^2\bar d^4\partial^{+(C-3)}\varphi^b)\,\partial^{+3}(\partial^{+(C-1)}\varphi^c\,\partial^{+(D)}\varphi^d)\Big]\ .
\end{align}
These two terms cancel if $f^{a}{}_{bcd}=-f^{a}{}_{dcb}$ and $C=D$. In a similar way,  it is not difficult to see that the $D-3/2$ term of \eqref{firstE} is also cancelled by the contributions that $\d^{2}\bar d^{4}$ acts on the ``second'' superfield $\partial^2\bar d^4\partial^{+(C-3)}\varphi^c$. 
It follows from \eqref{MMM-BECM} that 

\begin{align}
A\,=\,3\, ,\quad B\,=\,3\,,\quad M\,=\,-\,2\,,\quad C\,=\,D\,=\,1,\quad {\rm and}\quad f^{a}{}_{[bcd]}\ ,
\end{align}
which are the exponents for the BLG solution \eqref{BLGexp}.

When $m\ne 0$, we find for the most ``singular" term 

\begin{align}
\partial^{+2}\left\{\,\left[(\partial^{+(C-m)}\varphi^b\,\partial^{+(C+m)}\varphi^c-
\partial^{+(C-m+1)}\varphi^b\,\partial^{+(C+m-1)}\varphi^c
\,\right]\,\partial^2\bar d^4\partial^{+(D-3)}\varphi^d\,)\right\} \ .
\end{align}
which must cancel against \eqref{firstE}. However, these terms cannot cancel and no solution exists when $m\ne 0$.

\section{Even Ansatz: no BLG Solution}\label{App:even}
\setcounter{equation}{0}
For the even case, the commutator of the Hamiltonian with the boost is given by

\be 
[\,\deltab^{\rm even}_{{\cal P}^-}\,,\,\deltab^{^{\rm even}}_{{\cal J}^-}\,]\,\varphi^a_{}
~=~-\frac{1}{4}{\cal S}\,\frac{\partial^2}{\partial r\partial r'}\,\left({\mc F}\,{\cal O}^{{a,\,\rm even}
}_1+ {\mc G}\,{\cal O}^{a,\,\rm even}
_2\right)_{r=r'=0}+O(f^2)\ , \label{even-sol}
\ee
where 

\begin{align}
{\mc F}&\equiv~ (B^{}_{-1}-\frac{7}{2})\widehat {\mc U}^{}_1+(M_{-1}-C_{-1}-D_{-1}+3)\widehat {\mc U}^{}_2\ ,\cr
{\mc G}&\equiv~(C^{}_{-1}-1)\widehat {\mc U}^{}_3-(D_{-1}-1)\widehat {\mc U}^{}_4\ ,\label{Geven}
\end{align}
and

\bea
{\cal O}^{\rm even}_1&=& \sum_{\rm even}( K^{[r+r',1]}_{\alpha}-K^{[r+r',1]}_{\alpha+1}) -2\sum_{\rm odd} K^{[r,r']}_{\alpha+\frac{1}{2}}\ ,
\cr
{\cal O}^{\rm even}_2&=& \sum_{\rm odd}( K^{[1,r+r']}_{\alpha+\frac{1}{2}}-K^{[1,r+r']}_{\alpha+\frac{3}{2}}) +2\sum_{\rm even} K^{[r,r']}_{\alpha+1}\ .
\eea

We now search for solutions with integer-valued exponents. We show below that no such solutions exist for the even case, unlike the odd case.

 We first express the r.h.s. of \eqref{even-sol} in the base where $\alpha=-1$, and drop the subscripts. As done in the odd case, we only consider the terms of $\d^{2}\bar d^{4}$ acting on the same superfield.  
When $\d^{2}\bar d^{4}$ acts on the ``first'' superfield, we have

\begin{align}\label{firsteven}
- {\mc F} \frac{1}{\d^{+}} \Big( \frac{\d^{2}\bar d^{4}}{\d^{+5}}, \d^{+}(~,~) \Big) =
-(B-\frac{7}{2})\frac{1}{\d^{+}} \Big( \frac{\d^{2}\bar d^{4}}{\d^{+6}}, \d^{+}(~~,~~) \Big)
-(M-C-D+3)\frac{1}{\d^{+}} \Big( \frac{\d^{2}\bar d^{4}}{\d^{+5}}, (~~,~~) \Big)\ .
\end{align}
Notice that only the first term has the singular structure of $\frac{1}{\d^{+6}}$, which is different from the odd case whose singular structure lies on both terms.

The terms with $\d^{2}\bar d^{4}$ on the ``third'' superfield are given by the sum of the $\mc F$-terms and $\mc G$-terms

\begin{align}\label{evenFG}
&+{\mc F} \frac{1}{\d^{+}}\bigg[  -\Big(\d^{+},\frac{1}{\d^{+5}}(~,\d^{2}\bar d^{4}) \Big)
  -\d^{+}\Big(~,\frac{1}{\d^{+5}}(\d^{+},\frac{\d^{2}\bar d^{4}}{\d^{+}}) \Big)\\
&\qquad\qquad+\d^{+2}\Big(\frac{1}{\d^{+}},\frac{1}{\d^{+5}}(\d^{+2},\frac{\d^{2}\bar d^{4}}{\d^{+2}}) \Big)
+\d^{+3}\Big(\frac{1}{\d^{+2}},\frac{1}{\d^{+5}}(\d^{+3},\frac{\d^{2}\bar d^{4}}{\d^{+3}}) \Big)\bigg]\cr
&+2{\mc G}\frac{1}{\d^{+}}\bigg[+
\d^{+}\Big(~,\frac{1}{\d^{+6}}(\d^{+},\d^{2}\bar d^{4}) \Big)
+2\d^{+2}\Big(\frac{1}{\d^{+}},\frac{1}{\d^{+6}}(\d^{+2},\frac{\d^{2}\bar d^{4}}{\d^{+}}) \Big)+\d^{+3}\Big(\frac{1}{\d^{+2}},\frac{1}{\d^{+6}}(\d^{+3},\frac{\d^{2}\bar d^{4}}{\d^{+2}}) \Big)\bigg]\ ,\nn
\end{align}
which can also be written as

\begin{align}\label{FandGforeven}
&+{\mc F} \frac{1}{\d^{+}}\bigg[  -2 \Big(\d^{+},\frac{1}{\d^{+3}}(~,\frac{\d^{2}\bar d^{4}}{\d^{+2}}) \Big)
  -4 \Big(~,\frac{1}{\d^{+3}}(\d^{+},\frac{\d^{2}\bar d^{4}}{\d^{+2}}) \Big)
  -2\Big(\frac{1}{\d^{+}},\frac{1}{\d^{+3}}(\d^{+2},\frac{\d^{2}\bar d^{4}}{\d^{+2}}) \Big)\cr
&\qquad+\Big(\d^{+},\frac{1}{\d^{+2}}(~,\frac{\d^{2}\bar d^{4}}{\d^{+3}}) \Big)
+3\Big(~,\frac{1}{\d^{+2}}(\d^{+},\frac{\d^{2}\bar d^{4}}{\d^{+3}}) \Big)\cr
&\qquad+3\Big(\frac{1}{\d^{+}},\frac{1}{\d^{+2}}(\d^{+2},\frac{\d^{2}\bar d^{4}}{\d^{+3}}) \Big)
+\Big(\frac{1}{\d^{+2}},\frac{1}{\d^{+2}}(\d^{+3},\frac{\d^{2}\bar d^{4}}{\d^{+3}}) \Big)\bigg]\cr
&+2{\mc G}\bigg[\Big(~,\frac{1}{\d^{+4}}(\d^{+},\frac{\d^{2}\bar d^{4}}{\d^{+2}}) \Big)
+2\Big(\frac{1}{\d^{+}},\frac{1}{\d^{+4}}(\d^{+2},\frac{\d^{2}\bar d^{4}}{\d^{+2}}) \Big)
+\Big(\frac{1}{\d^{+2}},\frac{1}{\d^{+4}}(\d^{+3},\frac{\d^{2}\bar d^{4}}{\d^{+2}}) \Big)\bigg]\ .
\end{align}

In order to make sure that the most singular terms lie on $\varphi^{d}$ (not on $\varphi^{c}$), we assume that $C>D$. Then we follow singular terms with $(...\ ,\frac{1}{\d^{+n}} (...,...))$ structure.  
The most singular part reads

\begin{align*}
-2(M-2(C+D)+5)\frac{1}{\d^{+}}\bigg[ 
\Big( \d^{+},\frac{1}{\d^{+4}}(~,\frac{\d^{2}\dbar^{4}}{\d^{+2}})\Big)
+2\Big( ~,\frac{1}{\d^{+4}}(\d^{+},\frac{\d^{2}\dbar^{4}}{\d^{+2}})\Big)
+\Big( \frac{1}{\d^{+}},\frac{1}{\d^{+4}}(\d^{+2},\frac{\d^{2}\dbar^{4}}{\d^{+2}})\Big)
\bigg]\ .
\end{align*}
If $M+4>0$, then these terms must vanish by themselves, thus leading to the vanishing coefficient

\begin{align}\label{M2CD5}
M-2(C+D)+5 ~=~0 \ .
\end{align}
If $M+4\le 0$, on the other hand, these terms are no longer singular and thus they do not have to vanish by themselves.

 We then need to investigate terms along $\frac{\d^{2}\dbar^{4}}{\d^{+n}}$ singular structure.
Assuming that $M>-4$, we proceed with the terms of $\frac{\d^{2}\dbar^{4}}{\d^{+n}}$ singular structure.  After a little calculation with \eqref{M2CD5}, we find that the most singular terms are of $\frac{\d^{2}\dbar^{4}}{\d^{+3}}$,  and given by

\begin{align}\label{etwo}
&(B-\frac72)\frac{1}{\d^{+(A+1)}}\bigg( \d^{+3} [\d^{+(B-3)}\varphi^{b}\d^{+(C-M-2)}\varphi^{c}]\, \d^{2}
\dbar^{4}\d^{+(D-3)}\varphi^{d}\bigg) \cr
&+(M-C-3D+5)\frac{1}{\d^{+(A+1)}}\bigg( \d^{+3}[\d^{+(B-2)}\varphi^{b}\d^{+(C-M-3)}\varphi^{c}]
\d^{2}\dbar^{4}\d^{+(D-3)}\varphi^{d}\bigg)\ ,
\end{align}
where the first term comes from only the $\mc F$-terms while the last term from both the $\mc F$- and $\mc G$-terms in \eqref{FandGforeven}.  Since these terms are the most singular, they must either vanish by themselves or be canceled with \eqref{firsteven}. There are three possibilities for such cancelations: 
\begin{itemize}
\item The first term in \eqref{firsteven} cancels the first term of \eqref{etwo}, and the remaining term in \eqref{etwo} vanishes by itself due to either symmetry or a vanishing coefficient.  The necessary condition for this is to have the same powers of $\d^{+}$'s on the ``first'' and ``third'' superfields

\begin{align}\label{BD3}
D-3 = B-6 \quad \longrightarrow\quad B = D+3\ ,
\end{align}
and some symmetry in $f^{a}{}_{bcd}$ allowing the interchange of the indices $b$ and $d$. By comparing \eqref{etwo} with \eqref{firsteven}, we get

\begin{align}
&(B-\frac72)\frac{f^{a}{}_{bcd} }{\d^{+(A+1)}}\bigg( \d^{+3} [\d^{+D}\varphi^{b}\d^{+(C-M-2)}\varphi^{c}]\, \d^{2} \dbar^{4}\d^{+(D-3)}\varphi^{d}\bigg)  \cr
&= 
(B-\frac{7}{2})\frac{f^{a}{}_{dcb}}{\d^{+(A+1)}} \bigg( \d^{2}\dbar^{4} \d^{+(D-3)}\varphi^{d}\, \d^{+(1-M)}[\d^{+C}\varphi^{c}\d^{+D}\varphi^{b}] \bigg)\ ,
\end{align}
up to an overall sign, which implies that $M=-2$. It follows from \eqref{M2CD5} that 

\begin{align}\label{2CD3}
2\,(\,C\,+\,D\,)\,=\,3\ .
\end{align} 
Now we consider the remaining term, the second term in \eqref{etwo}, which must vanish either by the vanishing coefficient $M-C-3D+5$, or by the antisymmetry of $f^{a}{}_{[bc]d}$ requiring $B-2=C-M-3$. 
The vanishing coefficient, which gives $C+3D=3$, cannot lead to a solution because together with \eqref{2CD3}, it leads to $C=D$, which then contradicts the assumption $C>D$.  The other case with antisymmetry in $b$ and $c$ requires $B=C+1$, which leads to $4(D+1)=3$ to meet \eqref{BD3} and \eqref{2CD3}, but then this yields fractional powers. Hence, both cases do not yield  integer-valued solutions.

\item The second possibility comes from the observation that the first terms of \eqref{firsteven} and \eqref{etwo} have the same coefficients, which allows us to consider the possibility that these two terms can add up, to cancel against the second term of \eqref{etwo}. This also needs the condition \eqref{BD3}, but requires that

\begin{align} \label{2B7}
&(M-C-3D+5)
\frac{f^{a}{}_{bcd}}{\d^{+(A+1)}}\bigg( \d^{+3} [\d^{+(D+1)}\varphi^{b}\d^{+(C-M-3)}\varphi^{c}]\, \d^{2} \dbar^{4}\d^{+(D-3)}\varphi^{d}\bigg)  \cr
&= (2B-7)
\frac{f^{a}{}_{dcb}}{\d^{+(A+1)}} \bigg( \d^{2}\dbar^{4} \d^{+(D-3)}\varphi^{d}\, \d^{+(1-M)}[\d^{+C}\varphi^{c}\d^{+D}\varphi^{b}] \bigg)\ ,
\end{align}
up to an overall sign. Notice that the powers of $\d^{+}$'s on $\varphi^{b}$'s on both side are different, which makes us to impose further condition 

\begin{align}
D\,=\, C\,-\,M\,-\,2\ ,\quad f^{a}{}_{bcd}= -f^{a}{}_{cbd}\ 
\end{align}
so that

\begin{align}
&\frac{f^{a}{}_{bcd}}{\d^{+(A+1)}}\bigg( \d^{+3} [\d^{+(D+1)}\varphi^{b}\d^{+(C-M-3)}\varphi^{c}]\, \d^{2} \dbar^{4}\d^{+(D-3)}\varphi^{d}\bigg) \cr
& =\frac{f^{a}{}_{bcd}}{\d^{+(A+1)}}\bigg( \d^{+4} [\d^{+D}\varphi^{b}\d^{+(C-M-3)}\varphi^{c}]\, \d^{2} \dbar^{4}\d^{+(D-3)}\varphi^{d}\bigg) \ .
\end{align}
By comparing this to \eqref{2B7}, we find that $M=-3$, and thus $D=C+1$ which is in contradiction with the assumption $C>D$. 

\item 
The last possibility goes as follows. The first term in \eqref{etwo} vanishes due to a symmetry and the second term has the vanishing coefficient, regardless of the first term in \eqref{firsteven}, thus requiring 

\begin{align}\label{lastcond}
B-3=C-M-2\ , \ {\rm and}\  M-C-3D+5 =0 \quad \longrightarrow\quad 3D+B=6 \ ,
\end{align}
as well as \eqref{BD3}. These conditions then lead to another fractional solution $D=3/4$. Thus, this possibility does not yield integer-valued solutions.
\end{itemize}
We note that one might believe that there is another possibility that the second term of \eqref{etwo} can be canceled by the first term of \eqref{firsteven}, but this case cannot lead to integer valued solutions, so we neglect this possibility.\\

For $M+4<0$, \eqref{M2CD5} is not required, but the leading singular terms are, however, still of a similar form as  \eqref{etwo}

\begin{align}\label{etwo2}
&(B+2C-\frac{11}{2})\frac{1}{\d^{+(A+1)}}\bigg( \d^{+3} [\d^{+(B-3)}\varphi^{b}\d^{+(C-M-2)}\varphi^{c}]\, \d^{2}
\dbar^{4}\d^{+(D-3)}\varphi^{d}\bigg) \cr
&+(M-C-3D+5)\frac{1}{\d^{+(A+1)}}\bigg( \d^{+3}[\d^{+(B-2)}\varphi^{b}\d^{+(C-M-3)}\varphi^{c}]
\d^{2}\dbar^{4}\d^{+(D-3)}\varphi^{d}\bigg)\ .
\end{align}
Calculations for this case are similar to those for the odd case and there are no integer-valued solutions. 

Therefore, we have explored all possible cancelations for \eqref{etwo}, and showed that there are no integer-valued exponents that make \eqref{etwo} cancel out or vanish. Thus we conclude that there are no solutions  for the even case except the trivial one, \eqref{eventrivial}.


\setcounter{equation}{0}
\section{Bringing $\dab_{\epbar Q}^{int}\phi^a$ to the BLG form}\label{App:BLGform}

The conjugated ansatz (\ref{oddan}) in the odd case is
\bea
\dab_{\epbar\cQ}^{int}\phi^a=i\epbar_m\mc{S}^{-1}\frac{1}{3!}\eps^{i j k l}\left(
\frac{\p}{\p\eta^{i j k}}\frac{\p}{\p\zeta^l}\eta^m\cK_{-\half}^{a\,(0,\eta,\zeta)}
-\frac{\p}{\p\eta^i}\frac{\p}{\p\zeta^{j k l}}\eta^m\cK_{\half}^{a\,(0,\eta,\zeta)}\right)\Big|_{\eta=\zeta=0}.
\eea
Upon differentiating $\eta^m$ it becomes
\bea
\dab_{\epbar\cQ}^{int}\phi^a=-\frac{i}{3!}\epbar_m\Big(3\Psi_1^m-\Psi_2^m\Big),
\eea
where we defined
\bea
\Psi_1^m \equiv \eps^{m i j k}\frac{\p}{\p\eta^{i j}}\frac{\p}{\p\zeta^k}
\mc{S}^{-1}\cK_{-\half}^{a\,(0,\eta,\zeta)}\Big|_{\eta=\zeta=0}, \quad
\Psi_2^m \equiv \eps^{m i j k}\frac{\p}{\p\zeta^{i j k}}
\mc{S}^{-1}\cK_{\half}^{a\,(0,\eta,\zeta)}\Big|_{\eta=\zeta=0}.
\eea
With the BLG values for the exponents (\ref{BLGexp}), we have
\bea
\mc{S}^{-1}\cK_{-\half}^{a\, (0,\eta,\zeta)} &=& f^a{}_{b c d}
\frac{1}{\p^{+2}}\Big(\p^{+2}E_\eta\phi^b\cdot\p^{+}E_{-\eta}
(\p^{+}E_\zeta\phi^c\cdot\p^{+}E_{-\zeta}\phi^d)\Big),
\nn\\
\mc{S}^{-1}\cK_{\half}^{a\,(0,\eta,\zeta)} &=& f^a{}_{b c d}
\frac{1}{\p^{+}}\Big(\p^{+}E_\eta\phi^b\cdot\frac{1}{\p^{+}}E_{-\eta}
(\p^{+2}E_\zeta\phi^c\cdot\p^{+2}E_{-\zeta}\phi^d)\Big).
\eea
Differentiating and using the $[c d]$ antisymmetry, $f^a{}_{b c d}=-f^a{}_{b d c}$, we find
\bea
\Psi_1^m &=& 2f^a{}_{b c d}\eps^{m i j k}\frac{1}{\p^{+2}}\Big[
\dbar_{i j}\phi^b\cdot\p^{+}(\dbar_k\phi^c\cdot\p^{+}\phi^d)
+\p^{+2}\phi^b\cdot\frac{1}{\p^{+}}\dbar_{i j}(\dbar_k\phi^c\cdot\p^{+}\phi^d), \nn\\
&&\hspace{150pt}
-2\p^{+}\dbar_i\phi^b\cdot\dbar_j(\dbar_k\phi^c\cdot\p^{+}\phi^d)\Big] \nn\\
\Psi_2^m &=& 2f^a{}_{b c d}\eps^{m i j k}\frac{1}{\p^{+}}\Big\{
\p^{+}\phi^b\cdot\frac{1}{\p^{+}}\Big[
\frac{1}{\p^{+}}\dbar_{i j k}\phi^c\cdot\p^{+2}\phi^d
-3\dbar_{i j}\phi^c\cdot\p^{+}\dbar_k\phi^d\Big]\Big\}.
\eea
Using the $[b d]$ symmetry, $f^a{}_{b c d}=-f^a{}_{d c b}$, we can rewrite $\Psi_2^m$ as
\bea
\Psi_2^m &=& -2f^a{}_{b c d}\eps^{m i j k}\frac{1}{\p^{+}}\Big\{
\p^{+}\phi^b\cdot\frac{1}{\p^{+}}\Big[
\dbar_{i j k}\phi^c\cdot\p^{+}\phi^d
+3\dbar_{i j}\phi^c\cdot\p^{+}\dbar_k\phi^d\Big]\Big\}.
\eea
Similarly, using total antisymmetry $f^a{}_{b c d}=f^a{}_{[b c d]}$, we find that
\bea
\Psi_1^m=-\Psi_2^m
\eea
and therefore
\bea
\dab_{\epbar\cQ}^{int}\phi^a=-\frac{4i}{3}\epbar_m f^a{}_{b c d}\eps^{m i j k}
\frac{1}{\p^{+}}\Big\{
\p^{+}\phi^b\cdot\frac{1}{\p^{+}}\Big[
\dbar_{i j k}\phi^c\cdot\p^{+}\phi^d
+3\dbar_{i j}\phi^c\cdot\p^{+}\dbar_k\phi^d\Big]\Big\}.
\eea
Finally, using the following two forms of the inside-out constraint (\ref{inout})
\bea
\frac{1}{3!}\eps^{m i j k}\dbar_{i j k}\phi^c=(i\sqrt2\p^{+})d^m\phibar^c, \quad
\frac{1}{2}\eps^{m i j k}\dbar_{i j}\phi^c=d^{m k}\phibar^c
\eea
we arrive at the result given in (\ref{BLGsusy}).

\setcounter{equation}{0}
\section{``$C$-only'' projection of the Hamiltonian }\label{App:HisQF}

First, we rewrite equation (\ref{H1}) as
\bea
\label{H1XY}
H^{(1)}=\frac{8i}{2\sqrt2}\int d^3x \Big(2 X-i\sqrt2 Y\Big)+c.c.,
\eea
where, using the equivalence of $\int d^4\ta d^4\tabar$ to projecting with $d^{[4]}\dbar^{[4]}$, we defined
\bea
X &\equiv& f_{a b c d}d^{[4]}\dbar^{[4]}\Big\{ q^m\frac{\p}{\p^{+3}}\phi^a\cdot\p^{+}\phibar^b\cdot
\frac{1}{\p^{+}}(\p^{+}\dbar_m\phi^c\cdot\p^{+}\phibar^d) \Big\}{}_{\big|{\ta=\tabar=0}}, \nn\\[5pt]
Y &\equiv& f_{a b c d}d^{[4]}\dbar^{[4]}\Big\{ q^m\frac{\p}{\p^{+3}}\phi^a\cdot\p^{+}\phibar^b\cdot
\frac{1}{\p^{+}}(\dbar_{m n}\phi^c\cdot\p^{+}d^n\phibar^d) \Big\}{}_{\big|{\ta=\tabar=0}}.
\eea
Using that
\bea
d^{[4]}\dbar^{[4]} =\frac{1}{4!4!}\eps^{i j k l}\eps_{r s t u}d^{r s t u}\dbar_{i j k l}
\eea
and keeping only terms giving $C^{m n a}$ or $\Cbar_{m n}^a=\half\eps_{m n k l}C^{k l a}=(C^{m n a})^\ast$ upon projection,
\bea
\dbar_{m n}\phi^a{}_{\big|{\ta=\tabar=0}}=-i\sqrt2\,\Cbar_{m n}^a, \quad
d^{m n}\phibar^a{}_{\big|{\ta=\tabar=0}}=-i\sqrt2\,C^{m n a}
\eea
we find the following formulae
\bea
d^{[4]}\dbar^{[4]}\Big(q^m\phi,\;\phibar,\;\dbar_m\phi,\;\phibar\Big){}_{\big|\text{$C$-only}}
&=& 2i\sqrt2\Big(\p^{+}C^{m n},\;\Cbar_{i j},\;\Cbar_{m n},\;C^{i j}\Big), \nn\\
d^{[4]}\dbar^{[4]}\Big(q^m\phi,\;\phibar,\;\dbar_{m n}\phi,\;d^n\phibar){}_{\big|\text{$C$-only}}
&=& \nn\\
&&\hspace{-100pt}
-8\Big(\p^{+2}C^{m i},\;\Cbar_{i j},\;\Cbar_{m n},\;C^{n j}\Big) 
+16\Big(\p^{+}C^{m i},\;\Cbar_{i j},\;\p^{+}\Cbar_{m n},\;C^{n j}\Big) \nn\\ 
&&\hspace{-100pt}
-8\Big(\p^{+}C^{m n},\;\Cbar_{i j},\;\p^{+}\Cbar_{m n},\;C^{i j}\Big)
-4\Big(\p^{+}C^{m n},\;\Cbar_{i j},\;\Cbar_{m n},\;\p^{+}C^{i j}\Big).
\eea
Applying them to $X$ and $Y$, inserting an extra $\p^{+}/\p^{+}$ for $C^{m i a}$ in the first term in $Y$ and partially integrating, we find~\footnote{
Total $\p$ and $\p^{+}$ derivatives can be dropped under $\int d^3x$ in (\ref{H1XY}).
}
\bea
\mc{X} \equiv \left(\frac{i}{4\sqrt2}X+\frac{1}{8}Y\right){}_{\big|\text{$C$-only}}
&=& f_{a b c d}\Bigg\{
\frac{\p}{\p^{+2}}C^{m i a}\cdot\p^{+2}\Cbar_{i j}^b\cdot\frac{1}{\p^{+}}
\Big(\Cbar_{m n}^c\p^{+}C^{n j d}\Big) \nn\\
&&\hspace{-50pt}
+\frac{\p}{\p^{+2}}C^{m i a}\cdot\p^{+}\Cbar_{i j}^b\cdot\frac{1}{\p^{+}}\Big(
3\p^{+}\Cbar_{m n}^c\cdot\p^{+}C^{n j d}
+\Cbar_{m n}^c\p^{+2}C^{n j d}\Big) \nn\\
&&\hspace{-50pt}
-\frac{\p}{\p^{+2}}C^{m n a}\cdot\p^{+}\Cbar_{i j}^b\cdot\frac{1}{\p^{+}}\left(
\frac{3}{2}\p^{+}\Cbar_{m n}^c\cdot\p^{+}C^{i j d}
+\half\Cbar_{m n}^c\p^{+2}C^{i j d}\right) \Bigg\}. \qquad
\eea
Using the following identity
\bea
\label{CCCCid}
(C^{m i},\Cbar_{i j},\Cbar_{m n},C^{n j})
-\half(C^{m n},\Cbar_{i j},\Cbar_{m n},C^{i j})
=-(C^{m n},C^{i j},\Cbar_{m i},\Cbar_{n j}),
\eea
which follows from 
$\eps^{m n k[l}\eps^{i j r s]}(\Cbar_{k l},\Cbar_{i j},\Cbar_{m n},\Cbar_{r s})=0$ and 
$C^{m n a}\Cbar_{m n}^b=\Cbar_{m n}^a C^{m n b}$, we obtain
\bea
\mc{X} &=& f_{a b c d}\Big\{
\frac{\p}{\p^{+2}}C^{m i a}\cdot\p^{+2}\Cbar_{i j}^b\cdot\frac{1}{\p^{+}}
\Big(\Cbar_{m n}^c\p^{+}C^{n j d}\Big) \nn\\[5pt]
&&
-\frac{\p}{\p^{+2}}C^{m n a}\cdot\p^{+} C^{i j b}\cdot\frac{1}{\p^{+}}\Big[
3\p^{+}\Cbar_{m i}^c\cdot\p^{+}\Cbar_{n j}^d
+\Cbar_{m i}^c\p^{+2}\Cbar_{n j}^d\Big] \Big\}.
\eea
Using antisymmetry of $C$'s and $[c d]$ antisymmetry of $f_{a b c d}$, we find that the first term in the square bracket vanishes, whereas the other term can be written as a total derivative. Therefore,
\bea
\mc{X} &=& f_{a b c d}\Big\{
\frac{\p}{\p^{+2}}C^{m i a}\cdot\p^{+2}\Cbar_{i j}^b\cdot\frac{1}{\p^{+}}
\Big(\Cbar_{m n}^c\p^{+}C^{n j d}\Big) 
-\frac{\p}{\p^{+2}}C^{m n a}\cdot\p^{+} C^{i j b}\cdot\Cbar_{m i}^c\p^{+}\Cbar_{n j}^d \Big\}. \qquad
\eea
Partially integrating one $\p^{+}$ on $\Cbar_{i j}^b$ and using (\ref{CCCCid}) gives
\bea
\mc{X} &=& f_{a b c d}\Big\{
-\frac{\p}{\p^{+}}C^{m i a}\cdot\p^{+}\Cbar_{i j}^b\cdot\frac{1}{\p^{+}}
\Big(\Cbar_{m n}^c\p^{+}C^{n j d}\Big) \nn\\[5pt]
&&\hspace{60pt}
-\half\frac{\p}{\p^{+2}}C^{m n a}\cdot\p^{+}\Cbar_{i j}^b\cdot\Cbar_{m n}^c\p^{+}C^{i j d}\Big\}.
\eea
The second term vanishes thanks to $[b d]$ antisymmetry of $f_{a b c d}$. Partially integrating $\p^{+}$ on $\Cbar_{i j}^b$ in the remaining term, we find
\bea
\label{mcX}
\mc{X} &=& f_{a b c d}
\p C^{m i a}\cdot\Cbar_{i j}^b\frac{1}{\p^{+}}
\Big(\Cbar_{m n}^c\p^{+}C^{n j d}\Big),
\eea
where the other term vanished thanks to $[b c]$ antisymmetry of $f_{a b c d}$. Applying (\ref{CCCCid}) in the following form
\bea
(C^{m i},\Cbar_{i j},\Cbar_{m n},C^{n j})=\half(C^{i j},\Cbar_{i j},\Cbar_{m n},C^{m n})
-(\Cbar_{m i},C^{i j},\Cbar_{n j},C^{m n})
\eea
we obtain
\bea
\label{mcX1}
\mc{X}=f_{a b c d}\Big\{
\half\p C^{i j a}\cdot\Cbar_{i j}^b\frac{1}{\p^{+}}
\Big(\Cbar_{m n}^c\p^{+}C^{m n d}\Big)
-\p\Cbar_{m i}^a\cdot C^{i j b}\frac{1}{\p^{+}}
\Big(\Cbar_{n j}^c\p^{+}C^{m n d}\Big) \Big\}.
\eea
On another hand, complex conjugating $\mc{X}$ as given in (\ref{mcX}), we find
\bea
\label{mcX2}
\mc{X}^\ast=-f_{a b c d}\p\Cbar_{m i}^a\cdot C^{i j b}\frac{1}{\p^{+}}(\p^{+}\Cbar_{n j}^c\cdot C^{m n d}),
\eea
where we also used $[c d]$ antisymmetry of $f_{a b c d}$. Adding now (\ref{mcX1}) and (\ref{mcX2}), we find 
\bea
\mc{X}+\mc{X}^\ast=\half f_{a b c d}\,
\p C^{i j a}\cdot\Cbar_{i j}^b\frac{1}{\p^{+}}
\Big(\Cbar_{m n}^c\p^{+}C^{m n d}\Big),
\eea
where the other two terms cancel thanks to $[b d]$ antisymmetry of $f_{a b c d}$. It then follows that
\bea
H^{(1)}{}_{\big|\text{$C$-only}}
=32\int d^3x \Big(\mc{X}+\mc{X}^\ast\Big)=-16 f_{a b c d}\int d^3x
(\Cbar_{i j}^a\p C^{i j b})\frac{1}{\p^{+}}
(\Cbar_{m n}^c\p^{+}C^{m n d}),
\eea
where we used $[a b]$ antisymmetry of $f_{a b c d}$. This proves that (\ref{H1}) has the ``$C$-only'' part as given in (\ref{H1C}). On another hand, we have
\bea
d^{[4]}\dbar^{[4]}\Big\{(\phi^a\p\phi^b)\frac{1}{\p^{+}}(\phibar^c\p^{+}\phibar^d)\Big\}{}_{\big|\text{$C$-only}}
&=& \frac{6\cdot 6}{4!4!}\eps^{i j k l}\eps_{r s t u}
(\dbar_{i j}\phi^a\cdot\p\dbar_{k l}\phi^b)\frac{1}{\p^{+}}
(d^{r s}\phibar^c\cdot\p^{+}d^{t u}\phibar^d){}_{\big|\ta=\tabar=0} \nn\\
&=& \frac{(-i\sqrt2)^4}{16}\eps^{i j k l}\eps_{r s t u}
(\Cbar_{i j}^a\p\Cbar_{k l}^b)\frac{1}{\p^{+}}(C^{r s c}\p^{+}C^{t u d}) \nn\\
&=&(\Cbar_{i j}^a\p C^{i j b})\frac{1}{\p^{+}}
(\Cbar_{m n}^c\p^{+}C^{m n d})
\eea
and, as this result is real, this proves that (\ref{H1BN}) also has the ``$C$-only'' part as given in (\ref{H1C}).


\end{document}